\documentclass[a4paper,11pt]{article}

\usepackage{jheppub_mod} 
\usepackage[T1]{fontenc} 

\usepackage{ulem}
\usepackage{array}
\usepackage{booktabs}

\newcommand{\nnnl}{\nonumber\\}
\newcommand{\beq}{\begin{equation}}
\newcommand{\eeq}{\end{equation}}
\newcommand{\mpi}{M_\pi}
\newcommand{\metap}{M_{\eta'}}
\newcommand{\mD}{M_D}
\newcommand{\mK}{M_K}
\newcommand{\M}{\mathcal{M}}
\newcommand{\T}{\mathcal{T}}
\newcommand{\N}{\mathcal{N}}
\newcommand{\F}{\mathcal{F}}
\newcommand{\FF}{\mathrm{FF}}
\newcommand{\ol}{\bar}
\newcommand{\disc}{{\rm disc}\,}
\newcommand{\ra}{\rangle}
\newcommand{\rra}{\right\rangle}
\newcommand{\la}{\langle}
\newcommand{\lla}{\left\langle}

\newcommand{\diff}{\text{d}}
\renewcommand{\vec}[1]{\mathbf{#1}}

\title{\boldmath Dispersion-theoretical analysis of the $D^+\to K^-\pi^+\pi^+$ Dalitz plot}

\author{Franz Niecknig}
\author{and Bastian Kubis}

\affiliation{Helmholtz-Institut f\"ur Strahlen- und Kernphysik (Theorie) and\\
             Bethe Center for Theoretical Physics,
             Universit\"at Bonn,
             53115 Bonn, Germany}

\emailAdd{niecknig@hiskp.uni-bonn.de}
\emailAdd{kubis@hiskp.uni-bonn.de}

\abstract{
We study the Dalitz plot of the Cabibbo-favored charmed-meson decay $D^+\to K^-\pi^+\pi^+$
using dispersion theory.  The formalism respects all constraints from analyticity and unitarity,
and consistently describes final-state interactions between all three decay products.
We employ pion--pion and pion--kaon phase shifts as input, and fit the pertinent subtraction 
constants to Dalitz plot data by the CLEO and FOCUS collaborations.
Phase motions of resonant as well as nonresonant amplitudes are discussed, which should provide
crucial input for future studies of $CP$ violation in similar three-body charm decays.
}

\begin{document} 
\maketitle
\flushbottom

\section{Introduction}
\label{sec:intro}

Heavy-flavor three-body decays into light mesons provide a valuable source for Standard Model tests and beyond.
While they are driven, at short distances, by the weak interactions, their rich kinematic structure accessible in 
Dalitz plot distributions makes them a prime example for the application of modern tools of amplitude analysis~\cite{ATHOS}.  
A major motivation for the investigation of heavy-flavor decays is the study of $CP$ violation, which manifests itself 
in the appearance of (weak) phases and requires the interference of different amplitudes with, at the same time, 
different phases in the strong final-state interactions (see, e.g., Ref.~\cite{BigiSanda} for an in-depth overview).  
In contrast to (quasi-)two-body decays occurring at fixed total energies, three-body decays offer a resonance-rich 
environment with rapidly varying strong phases throughout the phase space available, which may strongly magnify the effects 
of $CP$ violation in certain parts of the Dalitz plot.

Obviously, in order to turn the search for potentially very small $CP$-violating phases in such complicated hadronic environments
into a precision instrument, it is inevitable to control the strong dynamics in the final state as accurately as possible,
in a model-independent fashion that, however, incorporates a maximum of theoretical and phenomenological constraints.
The traditional approach to model Dalitz plots in terms of isobars, i.e.\ a series of subsequent two-body decays, 
and describe the relevant line shapes in terms of Breit--Wigner (or Flatt\'e) functions, has clear limitations:
it fails to describe in particular the phase motion of the broad $S$-wave resonances such as
the $f_0(500)$ in pion--pion or the $K_0^*(800)$ in pion--kaon scattering
(see e.g.\ Refs.~\cite{GardnerMeissner,Johanna} in the context of heavy-flavor decays),
and neglects corrections beyond two-body rescattering in an unquantified manner.

It has therefore been advocated to employ the framework of dispersion theory for amplitude analyses~\cite{ATHOS},
which is built on unitarity, maximal analyticity, and crossing symmetry.
The dispersive framework adapted to study three-body decays was originally introduced by Khuri and Treiman 
for the decay $K\to 3\pi$~\cite{KhuriTreiman} and subsequently further developed~\cite{Bronzan,Aitchison:1965zz,Aitchison:1966,Pasquier:1968zz,Pasquier:1969dt}. The formalism has been resurrected in a modern form in Refs.~\cite{AnisovichLeutwyler,KWW}.
The Khuri--Treiman equations are based on elastic two-body unitarity and explicitly generate crossed-channel rescattering between the three final-state particles.
The equations are constructed by setting up dispersion relations for the crossed scattering processes, with a subsequent analytic continuation back into the decay region.
This continuation is performed along the lines of the continuation of the perturbative triangle graph and is extensively discussed in Ref.~\cite{Bronzan}.   

Khuri--Treiman equations have been successfully applied to various low-energy meson decays, like e.g.\ 
$\eta\to 3\pi$~\cite{AnisovichLeutwyler,KWW,Lanz,JLab:eta3pi}, $\omega/\phi\to 3\pi$~\cite{Niecknig,JLab:omega3pi},
or $\eta'\to\eta\pi\pi$~\cite{Schneider}.
In this work, we extend this formalism to three-body decays of open-charm mesons,
analyzing the Cabibbo-favored decays $D^+\to K^-\pi^+\pi^+/\bar{K}^0\pi^0\pi^+$. 
As input we solely rely on $\pi\pi$ and $\pi K$ phase-shift input. 
While these are not yet decay channels of major interest to study $CP$ violation,
the final-state interactions are going to be similar for others that are, 
such as the Cabibbo-suppressed decays $D\to 3\pi / \pi K\bar{K}$.
For the decays at hand, inelastic effects are small in large regions of the Dalitz plots, and therefore elastic unitarity provides a good approximation: the $\pi\pi$ channel allows for isospin 1 and 2 only, but no isoscalar components, 
which would necessitate a coupled-channel treatment, as a strong coupling to $K\bar{K}$ occurs.
The major inelasticities in the $\pi K$ channel are found to set in at the $\eta' K$ threshold~\cite{Jamin:2001zq,Edera:2005dk,Moussallam:2007qc}.

Thus with the high-statistics experimental data available~\cite{CLEO,FOCUS,Aitala:2005yh}, 
these decays provide a good test case to establish this dispersive framework in higher energy regions and set the path 
to Cabibbo-suppressed decays where traces of physics beyond the Standard Model may be searched for.
Besides, it allows for a further test of low-energy $\pi K$ and $\pi \pi$ dynamics as well as the importance of crossed-channel rescattering effects in three-body decays.
It may also provide an insight into scattering phase shifts at higher energies in the future.

The decay under consideration has been the subject of a number of previous theoretical publications, 
focusing on different issues raised by the experimental results.
One challenge is the proper treatment of the isospin 1/2 $S$-wave with the very broad, non-Breit--Wigner-shaped
$K_0^*(800)$ (or $\kappa$) resonance~\cite{DescotesGenon:2006uk}, and
the inclusion of two scalar resonances $K_0^*(800)$ and $K_0^*(1430)$ in a  way that conserves unitarity.
Furthermore, the width of the $K_0^*(1430)$ extracted from the experimental analyses in Refs.~\cite{CLEO,FOCUS} 
is found to be inconsistent with PDG values~\cite{PDG}.
In addition, the explicit comparison of the $\pi K$ partial-wave phases extracted from these decays~\cite{Aitala:2005yh,Link:2009ng} 
with $\pi K$ scattering results~\cite{Aston:1987ir} seems to indicate deviations from Watson's final-state theorem.

Ref.~\cite{Oller:2004xm} focuses on the isospin 1/2 $S$-wave final-state interactions, based on coupled-channel partial waves
for $K\pi$, $K\eta$, and $K\eta'$ constructed dispersively in Ref.~\cite{Jamin:2001zq}.  Decay and scattering data could be reconciled, 
although no three-body rescattering effects, isospin 3/2 components, or $\pi \pi$ channel were included.
Ref.~\cite{Boito:2009qd} similarly observes mutual consistency of $\pi K$ scattering and the $D$-meson decay, 
using related input to take 
two-body final-state interactions in the $\pi K$ isospin 1/2 $S$- and $P$-wave into account in terms of the 
corresponding scalar and vector form factors.
Furthermore, the short-distance weak interactions are described with the help of an effective Hamiltonian based on a factorization ansatz.
Again, weak repulsive partial waves (of isospin 3/2 and in the $\pi\pi$ system) as well as crossed-channel rescattering
are neglected.
We mention that similar approaches, using dispersively constructed form factors for two-body rescattering, 
but neglecting third-particle interactions,
have also been applied to $B\to K\pi\pi$ decays~\cite{ElBennich:2006yi,ElBennich:2009da}.

In Ref.~\cite{Magalhaes:2011sh}, a Faddeev-like equation is solved that builds up three-particle rescattering effects. The underlying two-particle $\pi K$ amplitudes are obtained form unitarized chiral perturbation theory fitted to experimental data.
The decay amplitude is simplified to include only the isospin 1/2 $S$-wave, aiming mainly at a study of the importance of rescattering effects and the reproduction of the experimental $S$-wave phases~\cite{Aitala:2005yh,Link:2009ng}.
The model for the weak vertex has subsequently been improved~\cite{Magalhaes:2015fva}.
Ref.~\cite{Guimaraes:2014cca} applies a similar approach with the addition of the isospin 3/2 $\pi K$ $S$-wave, 
but is still restricted to $S$-waves only. 
The only theoretical analysis known to us with all relevant partial waves, three-particle rescattering effects, and effects of the intermediate state $\ol{K}^0\pi^0\pi^+$ included, is Ref.~\cite{Nakamura:2015qga}.
The author performs a full Dalitz plot analysis on pseudo data, which we will later compare to.

The outline of this article is as follows.  
Section~\ref{sec:kinedecay} states some basic kinematical relations and shows both isospin and
partial-wave decomposition of the decay amplitude in question.
In Sec.~\ref{sec:dispersive}, we derive the coupled dispersive integral equations and discuss
how to solve these.
Numerical results are shown in Sec.~\ref{sec:numerics} and compared to experimental Dalitz plot
data by the CLEO~\cite{CLEO} and FOCUS~\cite{FOCUS} collaborations.
We conclude our study in Sec.~\ref{sec:conclusion}.
Some technical details are relegated to the appendices.

\section{Kinematics,  isospin decomposition, partial-wave expansion}
\label{sec:kinedecay}

\begin{figure}
\centering
\includegraphics[width=\linewidth]{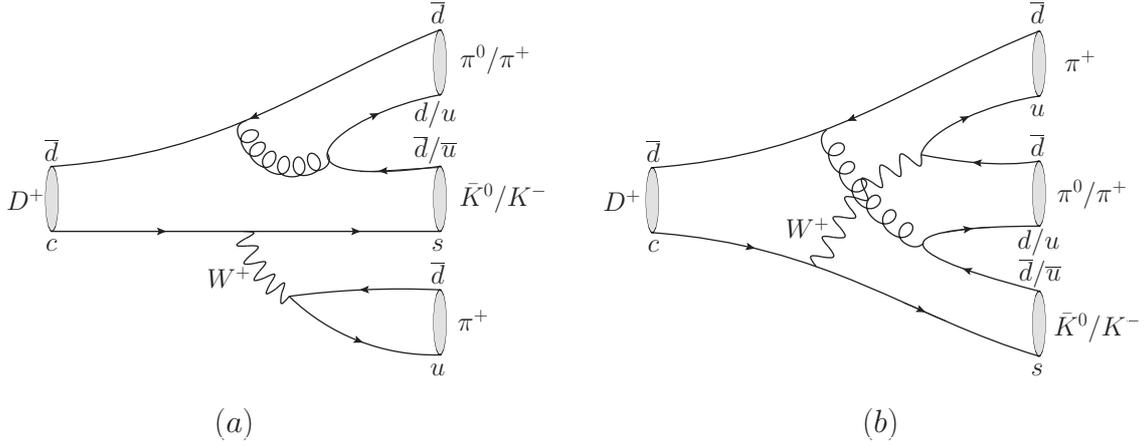}
\caption{Quark line diagrams of the $D^+\to K^-\pi^+\pi^+/\bar{K}^0\pi^0\pi^+$ decays: $W^+$ as a spectator  (a) and internal $W^+$ conversion (b).}
\label{fig:DKpipiq}
\end{figure}
The Mandelstam variables of the $D$-meson decay 
\begin{equation}
D^+(p_D)\to \bar K(p_K)\pi(p_1)\pi^+(p_2)
\end{equation}
are given by $s=(p_{D}-p_1)^2$, $t=(p_{D}-p_2)^2$, and $u=(p_{D}-p_{K})^2$.
The corresponding scattering angles $\theta$ in the (crossed) scattering processes are given by
\begin{align}
z_s &\equiv\cos\theta_s =\frac{s(t-u)-\Delta}{\kappa(s)}\,,
\quad
z_t\equiv\cos\theta_t=\frac{t(s-u)-\Delta}{\kappa(t)}\,,
\quad
z_u\equiv\cos\theta_u=\frac{t-s}{\kappa_u(u)}\,, \nnnl
\kappa(x) &=\lambda^{1/2}(x,\mK^2,\mpi^2)\lambda^{1/2}(x,\mD^2,\mpi^2)\,,
\quad
\kappa_u(u)=\lambda^{1/2}(u,\mD^2,\mK^2)\sqrt{1-\frac{4\mpi^2}{u}}\,,
\end{align} 
with $\Delta=\big(\mD^2-\mpi^2\big)\big(\mK^2-\mpi^2\big)$ 
and the  K\"all\'en function $\lambda(x,y,z)=x^2+y^2+z^2-2(xy+yz+xz)$.

We begin with the isospin and partial-wave decompositions of the decay amplitudes $\M_{-++}$ ($D^+\to K^-\pi^+\pi^+$) and $\M_{\ol{0}0+}$ ($D^+\to \bar{K}^0\pi^0\pi^+$).
We associate the isospin structure of the strong final-state current in Fig.~\ref{fig:DKpipiq} with the $D^+$ meson.
Since one $\bar{u}u$/$\bar{d}d$ pair is strongly produced, the associated isospin of the $D$ meson is given by $I=3/2,\,I_z=3/2$.
Thus the isospin decomposition of the respective (crossed) scattering processes reads
\begin{align}
s/t&\text{-channel} & u&\text{-channel}\nnnl
\M_{D^+\pi^0\rightarrow \bar{K}^0\pi^+} &=\sqrt{\frac{3}{5}} \F^{3/2}\,,           
 & \M_{D^+K^0\rightarrow \pi^0\pi^{+}}&=\frac{1}{2\sqrt{2}} \big(\F^2-\sqrt{3} \F^1\big)\,,\nnnl
\M_{D^+\pi^-\rightarrow K^-\pi^+}&=\sqrt{\frac{2}{15}} \F^{3/2}-\frac{1}{\sqrt{3}} \F^{1/2}\,, 
&\M_{D^+K^0\rightarrow \pi^+\pi^0}&=\frac{1}{2\sqrt{2}} \big(\F^2-\sqrt{3} \F^1\big)\,,\nnnl
\M_{D^+\pi^-\rightarrow \bar{K}^0\pi^0}&=\frac{2}{\sqrt{15}} \F^{3/2}+\frac{1}{\sqrt{6}}\F^{1/2}\,,
& \M_{D^+K^+\rightarrow \pi^+\pi^+}&=\F^2\,,   \label{eq:isoamp}
\end{align}
where  $\F^I$ denotes the amplitude with definite isospin $I$. The decay amplitudes are given by
\begin{align}
\M_{-++}(s,t,u)&=\M_{D^+\pi^-\rightarrow K^-\pi^+}(s,t,u)+\M_{D^+\pi^-\rightarrow K^-\pi^+}(t,s,u)+\M_{D^+K^+\rightarrow \pi^+\pi^+}(s,t,u), \nnnl
\M_{\ol{0}0+}(s,t,u)&=\M_{D^+\pi^0\rightarrow \bar{K}^0\pi^+}(s,t,u)+\M_{D^+K^0\rightarrow \pi^0\pi^{+}}(s,t,u)+\M_{D^+\pi^-\rightarrow \bar{K}^0\pi^0}(s,t,u)\,.
\end{align}
We can write down a symmetrized partial-wave expansion simultaneously in $s$-, $t$-, and $u$-channels
(the precise relation of which to proper partial waves in a single channel will be discussed below).
With the  expansion in partial-wave amplitudes 
truncated at the $D$-wave for $\pi K$ final states and the $P$-wave for $\pi \pi$, we obtain 
\begin{align}
\M_{-++}(s,t,u)&=\F^2_0(u)
+\bigg\{ \frac{1}{\sqrt{3}}\F^{1/2}_0(s)
-\sqrt{\frac{2}{15}}\F^{3/2}_0(s) \nnnl
&+\big[s(t-u)-\Delta\big]\bigg(\frac{1}{\sqrt{3}}\F^{1/2}_1(s)-\sqrt{\frac{2}{15}}\F^{3/2}_1(s)\bigg)\nnnl
&+\frac{1}{2}\big[3\big(s(t-u)-\Delta\big)^2-\kappa^2(s)\big]
\bigg(\frac{1}{\sqrt{3}}\F^{1/2}_2(s)-\sqrt{\frac{2}{15}}\F^{3/2}_2(s)\bigg) + (s \leftrightarrow t) \bigg\}\,,\nnnl
\M_{\ol{0}0+}(s,t,u)&=\frac{1}{2\sqrt{2}}\big(-\F^2_0(u)+\sqrt{3}(t-s)\,\F^1_1(u)\big) 
+\sqrt{\frac{3}{5}} \F^{3/2}_0(s) \nnnl
&+\sqrt{\frac{3}{5}}\big[s(t-u)-\Delta\big]\, \F^{3/2}_1(s) 
+\frac{\sqrt{3}}{2\sqrt{5}}\big[3\big(s(t-u)-\Delta\big)^2-\kappa^2(s)\big]\,\F^{3/2}_2(s) \nnnl
&-\bigg(\frac{2}{\sqrt{15}}\F^{3/2}_0(t)+\frac{1}{\sqrt{6}}\F^{1/2}_0(t) \bigg)
\nnnl
&
-\big[t(s-u)-\Delta\big]\bigg(\frac{2}{\sqrt{15}}\F^{3/2}_1(t)+\frac{1}{\sqrt{6}}\F^{1/2}_1(t)\bigg) \nnnl
&-\frac{1}{2}\big[3\big(t(s-u)-\Delta\big)^2-\kappa^2(t)\big]\bigg(\frac{2}{\sqrt{15}}\F^{3/2}_2(t)+\frac{1}{\sqrt{6}}\F^{1/2}_2(t)\bigg)
\,,\label{eq:Fullamp}
\end{align}
where the single-variable amplitudes $\F^I_L$ have definite isospin $I$ and angular momentum $L$ in the channel
associated with the Mandelstam variable featuring as their argument.
Note that the inclusion of $D$-waves is somewhat heuristic: in order to rigorously
prove the symmetrized decomposition~\eqref{eq:Fullamp} in the spirit of the so-called 
reconstruction theorem~\cite{Stern,Knecht,KWW,Anant,Zdrahal,StofferDiss}, one needs to include a subtraction
polynomial of higher order (i.e., a larger number of unknown parameters) than what we will allow for below. 
We mainly want to retain the $\pi K$ $D$-wave to test the effect of the $K_2^*(1430)$ resonance, which is kinematically
accessible in the decay phase space.  The way we implement this approximately will be discussed in Sec.~\ref{subsec:subtraction}.

\section{Dispersive formalism}\label{sec:dispersive}

\begin{figure}
\centering
\includegraphics[width= 0.5\linewidth ]{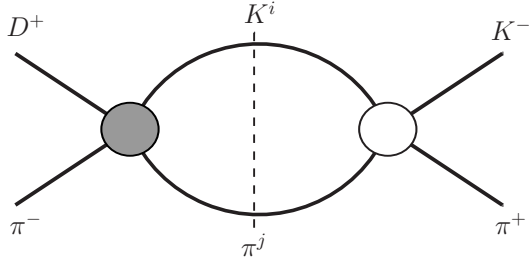}
\caption{The associated $s$-channel scattering diagram $D^+\pi^-\to K^-\pi^+$ via the intermediate states $K^i\pi^j$ . The gray vertex stands for the crossed decay amplitude $D^+\pi^-\to K^i\pi^j$ 
denoted by $\M_{ij+}$ and the white vertex the $K^i\pi^j\to  K^-\pi^+$ scattering amplitude denoted by $T^{ij,-+}$.
The dashed line gives the contribution to the discontinuity~\cite{Cutkosky}. The other channels follow analogously. }
\label{fig:Cutting}
\end{figure}
 
\subsection{Unitarity and Omn\`es solution}\label{subsec:Omnes}

We begin with the dispersive treatment of the associated scattering processes 
linked to the decay by crossing symmetry,
$D^+ \ol{\pi}\to K \pi$ and $ D^+ \bar{K}\ \to \pi \pi$.
The $D$-meson mass is artificially set to $\mD<\mK+2\mpi$ such that the corresponding decay is kinematically forbidden.
The simpler analytic structure of these scattering processes can be exploited to construct dispersion relations for the single-variable amplitudes valid for $s,t>(\mD+\mpi)^2$ and $u>(\mD+\mK)^2$, respectively.
The analytic continuation back to the physical $D$-meson mass as well as into the kinematic region 
$(\mK+\mpi)^2<s,t<(\mD-\mpi)^2$, $4\mpi^2<u<(\mD-\mK)^2$ yields the anticipated decay amplitudes~\cite{Bronzan}. 

We demonstrate the framework for the example of the $s$-channel processes; 
the $t$- and $u$-channel amplitudes are constructed analogously.
Elastic unitarity gives for the discontinuity (see Fig.~\ref{fig:Cutting} for $\M_{-++}$) 
\begin{align}
\disc\M_{-++}(s,z_s)&=\frac{i}{2}\int\frac{\diff^4l}{(2\pi)^2}\sum_{(i,j)}\M_{ij+}(s,z_s') \T^{ij,-+*}(s,z_s'')\delta\big(l^2-M_i^2\big)\,\delta\big((q-l)^2-M_j^2\big) \,,\nnnl
\disc\M_{\ol{0}0+}(s,z_s)&=\frac{i}{2}\int\frac{\diff^4l}{(2\pi)^2}\sum_{(i,j)}\M_{ij+}(s,z_s') \T^{ij,\ol{0}0*}(s,z_s'')\delta\big(l^2-M_i^2\big)\,\delta\big((q-l)^2-M_j^2\big)\label{eq:optthm}\,,
\end{align}
where  $\T^{ij,-+}(x,z_x)$  ($K^i\pi^j\to K^-\pi^+$) and $\T^{ij,\ol{0}0}(x,z_x)$ ($K^i\pi^j \to \bar{K}^0\pi^0$) are the inter\-mediate-to-final-state scattering amplitudes. $q=p_K+p_2= (\sqrt{s},\vec{0})$ defines the center-of-mass frame, in which $z_s'=\cos\theta'_s$, 
the cosine of the angle between initial and intermediate states, and $z_s''=\cos\theta''_s$, the cosine of the angle between intermediate and final state, are evaluated.
The intermediate-state summation runs over the tuple $(i,j)\in\{(-,+),(\ol{0},0)\}$.
The partial-wave decompositions for the  $\pi K$ ($\pi\pi$) amplitudes $\T^{ij,kl}$ and full decay amplitudes $\M_{ijk}$ read
\begin{align}
\T^{ij,kl}(s,z_s)&=\sum_{I,L}   a_{I,L}^{ij,kl}\,P_L(z_s)\,  t_L^I(s)\,, \nnnl
\M_{ijk}(s,z_s)&=\sum_{I,L} a_{I,L}^{ijk}\,P_L(z_s)\,  f^I_L(s)\,, \label{eq:partialwaves}
\end{align}
where the sum runs over isospin and angular momentum components $I$ and $L$. Furthermore we use the Clebsch--Gordan coefficients $a_ {I,L}$, Legendre polynomials $P_L(z)$, and the corresponding partial waves  
$t_L^I(s)$ and $f_L^I(s)$.\footnote{Note that in contrast to the definition of the single-variable amplitudes
in Eq.~\eqref{eq:Fullamp}, we have not defined the partial waves in Eq.~\eqref{eq:partialwaves} to be free of
kinematical zeros.  This is independent of the singularities these partial waves display at the corresponding
pseudo-thresholds or upper limits of the physical decay region, $s=(M_D-M_\pi)^2$ or $u=(M_D-M_K)^2$, which
are well understood, see e.g.\ Ref.~\cite{Niecknig} or the discussion in Ref.~\cite{GKR} in a perturbative
context.}
Exploiting the unitarity relation for elastic $\pi K$ and $\pi\pi$ scattering we obtain the following partial-wave unitarity relations 
\beq\label{eq:Watson}
 \disc f_L^I(s) = 2i\,f_L^I(s)\sin\delta_L^I(s)e^{-i\delta_L^I(s)}\theta\big(s-s_{\rm th}\big)\,,
\eeq
where  $\delta_L^I(s)$ denotes the elastic final-state scattering phase shift. 
The thresholds in the different channels are
$s_{\rm th} = t_{\rm th} = (\mK+\mpi)^2$  for $\pi K$ and $u_{\rm th} = 4\mpi^2$ for $\pi\pi$ scattering, respectively.
Since the discontinuity of $f_L^I$ and the according single-variable amplitude $\kappa^L\,\F_L^I$ coincide on the right-hand cut, we have 
\begin{align}
\disc f_L^I(s)&=\kappa^L(s)\,\disc \F_L^I(s)\, \nnnl
\Rightarrow \quad f_L^I(s)&=\kappa^L(s)\,\big(\F_L^I(s)+\hat\F_L^I(s)\big)\,,\label{eq:inhomodef}
\end{align}
where we have introduced the inhomogeneities $\hat\F_L^I(s)$ that are free of discontinuities on the right-hand cut by construction.
They incorporate the left-hand cut contributions and will be further discussed in Sec.~\ref{subsec:inhomogeneities}.
From Eqs.~\eqref{eq:inhomodef} and \eqref{eq:Watson} we obtain
\beq\label{eq:unrel}
 \disc\F_L^I(s) = 2i\,\bigl(\F_L^I(s)+\hat\F_L^I(s)\bigr)\,\theta(s-s_{\rm th})\sin\delta_L^I(s)e^{-i\delta_L^I(s)}\,,
\eeq
which has the form of an inhomogeneous Hilbert-type equation.
The homogeneous solution $\hat\F_L^I(s)=0$ is given by the so-called Omn\`es function $\Omega_L^I(s)$~\cite{Omnes} times an analytic function $P_L^I(s)$,
\beq
\F_L^I(s)=P_L^I(s)\Omega_L^I(s) \,, \quad  \Omega_L^I(s)=\exp\biggl\{\frac{s}{\pi}\int_{s_{\rm th}}^\infty \diff s'\frac{\delta_L^I(s')}{s'(s'-s)}\biggr\}\,.
\eeq
The inhomogeneous solution is obtained by a product ansatz  
\beq\label{eq:Khuri}
\F_L^I(s)=\Omega_L^I(s)\biggr\{P_L^I(s) + \frac{s^n}{\pi}\int_{s_{\rm th}}^{\infty}\frac{\diff s'}{s'^n}\frac{\sin\delta_L^I(s')\hat\F_L^I(s')}{|\Omega_L^I(s')|(s'-s)}\biggr\}\,,
\eeq
where $P_L^I(s)$ is now a polynomial of order $n-1$, and the number of subtractions $n$ is chosen such that the convergence of the dispersion integral is guaranteed.

As our approach relies on elastic unitarity (see Ref.~\cite{Guo:2015kla} for a generalization of the Khuri--Treiman formalism 
to coupled channels), 
the formalism breaks down when inelastic channels become important. 
We assume that Watson's theorem~\cite{Watson:1954uc} is a good approximation up to the $\eta' K$ threshold in the $\pi K$ channel.
Inelastic effects in the prominent $\pi K$  $S$-wave systems are found to become sizable above the  $\eta' K$ threshold~\cite{Jamin:2001zq,Edera:2005dk,Moussallam:2007qc}. 
The main inelastic contributions in the isospin 1/2 $P$-wave come from the $\pi K^*$ and  $\rho K$ channels, which become noticeable in the energy region where they couple to $K^*(1410)$ and $K^*(1690)$~\cite{Moussallam:2007qc}.
In all exotic partial waves, i.e.\ the isospin 2 $\pi\pi$ system as well as
$I=3/2$ $\pi K$ partial waves, inelastic effects are assumed to be negligible.

\subsection{Inhomogeneities}\label{subsec:inhomogeneities}

\begin{sloppypar}
With the scattering phase shifts given as fixed input, 
the only quantities left in the dispersion integrals Eqs.~\eqref{eq:Khuri} are the 
inhomogeneities $\hat{\F}^I_L$, which are determined as the projections of the crossed-channel amplitudes 
onto the considered channel. 
They can be re-expressed in terms of the single-variable amplitudes $\F_L^I(x)$, such that we obtain integral equations that can be solved for the $\F_L^I(x)$. 
With the aid of Eq.~\eqref{eq:inhomodef} we find
\begin{align}
f_L^I(x)&=\frac{2L+1}{2a_{I,L}^{ijk}}\int \diff z_x\M_{ijk}^{I_x}(x,z_x)P_L(z_x)=\kappa^L(x)\big(\F_L^I(x)+\hat{\F}_L^I(x)\big)\nnnl
\Rightarrow \quad \hat{\F}_L^I(x)&=\frac{2L+1}{2a_{I,L}^{ijk}\,\kappa^L(x)}\int_{-1}^1 \diff z_x\M_{ijk}^{I_x}(x,z_x)P_L(z_x)-\F_L^I(x)\,,\label{eq:inhomodet}	
\end{align}
where $\M_{ijk}^{I_x}(x,z_x)$ denotes the projection of the full decay amplitude $\M_{ijk}(x,z_x)$  onto isospin $I_x$ eigenfunctions in the $x$-channel.
One term of the projection integral over $\M_{ijk}^{I_x}(x,z_x)$ is always $\F_L^I(x)$, such that the right-hand-cut discontinuity is canceled. The 
inhomogeneities are then indeed free of discontinuities on the right-hand cut as anticipated. The resulting inhomogeneities are given in Appendix~\ref{app:inhomo}.
\end{sloppypar}

The interpretation of Eq.~\eqref{eq:inhomodet} as an angular integration is valid in the scattering region and needs to be analytically continued into the unphysical and decay regions. Performing the angular integration naively in the decay region results in crossing the unitarity cut.
The prescription on how to perform the continuation has been extensively discussed in Ref.~\cite{Bronzan}, motivated by the continuation of the (perturbative) triangle graph into the decay region.
It ultimately leads to the prescription $M_D^2\to M_D^2+i\epsilon$, which allows one to derive an integration path that avoids the unitarity cut.

\subsection{Number of subtraction constants}\label{subsec:subtraction}
The minimal number of subtractions needed is dictated by the asymptotic behavior of the integrands in Eqs.~\eqref{eq:Khuri}.
The decay amplitude and thus the inhomogeneities are assumed to grow at most linearly asymptotically, loosely based on the Froissart bound~\cite{Froissart}. 
Assuming the phase shifts to approach constant values $\delta_L^I(\infty)$ 
for large energies,
the Omn\`es functions $\Omega_L^I(x)$ behave like $\propto x^{-\delta_L^I(\infty)/\pi}$ asymptotically. With the following assumption for the phase shifts $\delta_L^I$ at high energies:
\begin{align}
\lim_{x\to\infty}\delta_0^{1/2}(x)&=2\pi\,, 	& \lim_{x\to\infty}\delta_1^{1/2}(x)&=\pi\,, 	 & \lim_{x\to\infty}\delta_2^{1/2}(x)&=\pi\,,\nnnl
\lim_{x\to\infty}\delta_0^{3/2}(x)&=0\,,		& \lim_{x\to\infty}\delta_1^{3/2}(x)&=0\,, 	& \lim_{x\to\infty}\delta_2^{3/2}(x)&=0\,, 	 \nnnl
\lim_{x\to\infty}\delta_0^2(x)&=0\,, 		&\lim_{x\to\infty}\delta_1^1(x)&=\pi\,,		 &
\end{align}
we need two subtractions for $\F_0^2$, $\F_1^1$, and $\F_0^{3/2}$, four subtractions for  $\F_0^{1/2}$, and one subtraction for $\F_1^{1/2}$ to obtain convergent dispersion integrals.
Note that the difference in the number of subtractions for $\F_1^1$ and $\F_1^{1/2}$, 
despite identical phase asymptotics, is due to the different kinematic
prefactors for $P$-waves with equal and unequal masses, see Eq.~\eqref{eq:Fullamp}.
$\F_1^{3/2}$ needs no subtraction, but as the $\pi K$ isospin 3/2 $P$-wave phase shift is very small and assumed to vanish at high energies, we neglect it altogether.  Similarly, also the $I=3/2$ $D$-wave is put to zero.

The inclusion of the $D$-wave $\F_2^{1/2}$ is delicate.  Formally it requires no subtractions, but the 
kinematical pre-function corresponding to the $L=2$ Legendre polynomial, multiplied with the required momentum factors
to make it free of kinematical singularities, see Eq.~\eqref{eq:Fullamp},
 violates the assumed high-energy behavior of the decay amplitude and thus of all inhomogeneities. 
Therefore we will follow a ``hybrid approach'' for the $D$-wave:
we will only consider the projections of $S$- and $P$-waves of other channels in order to generate
the $D$-wave inhomogeneity, but will exclude $D$-wave projections, thus eschewing the need for further 
subtractions.  This is loosely motivated by analogous observations in low-energy processes calculated
in chiral perturbation theory, where higher partial waves are dominated by crossed-channel loop diagrams
that correspond to low partial waves in those crossed channels.  

In total we have eleven subtraction constants.
However, the resulting representations of the decay amplitudes Eqs.~\eqref{eq:Fullamp} are not unique due to the linear dependence of the Mandelstam variables $s$, $t$, and $u$:
One can construct polynomial contributions to the single-variable amplitudes that leave the complete decay amplitudes $\M_{-++}(s,t,u)$ and $\M_{\ol{0}0+}(s,t,u)$ invariant; this is obvious in a standard dispersive representation, however slightly less trivial to demonstrate in the Omn\`es
representations discussed above~\cite{Leutwyler}.
The polynomial coefficients can be tuned such that a maximal number of subtraction constants is eliminated to obtain a linearly independent set.
These polynomials span the so-called invariance group of the decay amplitudes. 
Details are discussed in Appendix~\ref{app:Invariance}.
We choose to eliminate the subtraction constants in the nonresonant $I=3/2$ $\pi K$ and $I=2$ $\pi\pi$ $S$-waves,
the rationale being solely to retain them in presumably large, resonant partial waves.
This leaves seven linearly independent complex subtraction constants, 
\begin{align}
\F_0^2(u)&=\Omega^2_0(u)\frac{u^2}{\pi}\int_{u_{\rm th}}^\infty \frac{\diff u'}{u'^2}
\frac{\hat{\F}_0^2(u')\sin\delta^2_0(u')}{\big|\Omega^2_0(u')\big|(u'-u)} \,,\nnnl
\F_1^1(u)&=\Omega^1_1(u)\bigg\{c_0+c_1u+\frac{u^2}{\pi}\int_{u_{\rm th}}^\infty\frac{\diff u'}{u'^2}
\frac{\hat{\F}_1^1(u')\sin\delta^1_1(u')}{\big|\Omega^1_1(u')\big|(u'-u)}\bigg\}\,,\nnnl
\F_0^{1/2}(s)&=\Omega^{1/2}_0(s)\bigg\{c_2+c_3s+c_4s^2+c_5s^3+\frac{s^4}{\pi}\int_{s_{\rm th}}^\infty\frac{\diff s'}{s'^4}
\frac{\hat{\F}_0^{1/2}(s')\sin\delta^{1/2}_0(s')}{\big|\Omega^{1/2}_0(s')\big|(s'-s)}\bigg\}\,,\nnnl  
\F_0^{3/2}(s)&=\Omega^{3/2}_0(s)\bigg\{\frac{s^2}{\pi}\int_{s_{\rm th}}^\infty\frac{\diff s'}{s'^2}
\frac{\hat{\F}_0^{3/2}(s')\sin\delta^{3/2}_0(s')}{\big|\Omega^{3/2}_0(s')\big|(s'-s)}\bigg\}\,,\nnnl
\F_1^{1/2}(s)&=\Omega^{1/2}_1(s)\bigg\{c_6+\frac{s}{\pi}\int_{s_{\rm th}}^\infty\frac{\diff s'}{s'}
\frac{\hat{\F}_1^{1/2}(s')\sin\delta^{1/2}_1(s')}{\big|\Omega^{1/2}_1(s')\big|(s'-s)}\bigg\}\,,\nnnl        
\F_2^{1/2}(s)&=\Omega^{1/2}_2(s)\frac{1}{\pi}\int_{s_{\rm th}}^\infty\diff s'
\frac{\hat{\F}_2^{1/2}(s')\sin\delta^{1/2}_2(s')}{\big|\Omega^{1/2}_2(s')\big|(s'-s)}\,.\label{eq:fulleq}
\end{align}
The subtraction constants cannot be determined in the framework of dispersion theory and have to be obtained either by matching to a more fundamental dynamical theory, or, as in this work, by a fit to experimental data.
The solution space of the coupled system Eq.~\eqref{eq:fulleq} has thus dimension seven, corresponding to the seven complex subtraction constants. 
Since the equations depend linearly on the subtraction constants, it is convenient to choose seven independent basis
sets and solve the equations for each of these sets.
We call those solutions basis functions.
In particular, we choose for the $i$th basis function $\M_i(s,t,u)$ the set of subtraction constants $c_j=\delta_{ij}$ with $i,j=0\dots6$.
The full solution $\M(s,t,u)$ is then obtained by
\beq
\M(s,t,u)=\sum_{i} c_i \M_i(s,t,u)\,.
\eeq
The basis functions are entirely determined by the phase shift input, as well as the masses of all particles involved 
(taken from Ref.~\cite{PDG}).
The phase shifts are obtained from solutions of the $\pi\pi$ Roy equations by
both the Bern~\cite{ACGL,Caprini:2011ky,CapriniWIP} and the Madrid--Krak\'ow~\cite{Pelaez} groups,
as well as the Roy--Steiner equations for $\pi K$ scattering solved by the Orsay group~\cite{piK}.

\subsection{Solution strategy}\label{subsec:solution}

In this section we discuss different solution strategies of the Khuri--Treiman-type equations~\eqref{eq:fulleq}, their issues, and present our new solution strategy.

The standard solution strategy for the linear coupled double integral equations~\eqref{eq:fulleq} has been an iteration procedure as performed for example in Refs.~\cite{Lanz,Niecknig,Schneider} or numerically faster with the introduction of integral kernels in Ref.~\cite{Descotes-Genon:2014tla}.
Starting with an arbitrary input for the single-variable amplitudes (e.g.\ just the Omn\`es functions), the inhomogeneities are evaluated; with these the dispersion integrals are determined to obtain a new set of single-variable amplitudes.
This cycle is repeated until satisfactory convergence is reached.
Unfortunately, the convergence of this iterative procedure is not always guaranteed, 
depending on the mass of the decaying particle and the number of subtractions:
for larger decay masses and more subtractions applied, the corrections in each iteration step can be too large to reach the fixed-point solution.  We find this to be the case in the $D$-meson decays considered here. 

This necessitates a different solution strategy.
Since the set of integral equations is linear in the single-variable amplitudes it is convenient to set it up in the form of a matrix equation instead. Provided that the matrix is invertible a unique solution exists.
One such inversion strategy is known as the Pasquier inversion~\cite{Pasquier:1968zz}
(see Ref.~\cite{JLab:methods} for a recent comparison of Pasquier inversion and iterative solution),
where a method to reduce the double integral equation to a single integral equation is introduced.
The procedure involves the deformation of the integral contours of both integrals, 
allowing one to interchange the order of integrations such that a unique kernel function is obtained.
The  coupled single integral equations thus obtained do allow for a direct solution via matrix inversion.

We will follow a slightly modified strategy, 
constructing a matrix equation without performing a Pasquier inversion. 
In this context it is beneficial to solve for the inhomogeneities instead of the single-variable amplitudes,
the advantage being that the inhomogeneities need to be evaluated only on the right-hand-cuts. 
The single-variable amplitudes themselves can be obtained in the whole complex plain in a straightforward manner 
by performing the dispersion integral over the inhomogeneities once.

To illustrate the solution strategy we limit ourselves to one hypothetical inhomogeneity equation without any loss of generality,
\beq \label{eq:anghyp}
\hat{\F}_L(s)=\frac{1}{2}\int_{-1}^1 \diff z_s\, z_s^m \, \F\big(t(s,z_s)\big)\,, 
\eeq
and focus on the functions $\tilde{\F}(s)\equiv\hat{\F}_L(s)\kappa^{2L+1}(s)$ that are free of singularities at the 
pseudo-threshold or upper limit of the kinematically accessible decay region (which is a zero in $\kappa(s)$).
Inserting Eq.~\eqref{eq:Khuri} into Eq.~\eqref{eq:anghyp} yields 
\begin{align}
\tilde{\F}(s)&=\frac{\kappa^{2L+1}(s)}{2}\int_{-1}^1 \diff z_s \, z_s^m\, \Omega\big(t(s,z_s)\big)\biggl\{P\big(t(s,z_s)\big) \nnnl
&\quad +\frac{t(s,z_s)^n}{\pi}\int_{s_{\rm th}}^\infty \frac{\diff x}{x^n}
\frac{\tilde{\F}(x)\sin\delta(x)}{|\Omega(x)|\kappa^{2L+1}(x)\big(x-t(s,z_s)\big)} \biggr\} \nnnl
&\equiv A(s)+\frac{1}{\pi}\int_{s_{\rm th}}^\infty\tilde{\F}(x)K(s,x)\diff x\,. \label{eq:Fint}
\end{align}
The function $A(s)$ contains the dependence on the subtraction polynomial, while the integration kernel $K(s,x)$ 
is independent of any subtraction constants:
\begin{align}
A(s)&= \frac{\kappa^{2L+1}(s)}{2}\int_{-1}^1 \diff z_s\,z_s^m\,P\big(t(s,z_s)\big)\,\Omega\big(t(s,z_s)\big)\,, \nnnl
K(s,x)&= \kappa^{2L+1}(s)\frac{\sin\delta(x)}{x^n|\Omega(x)|\kappa^{2L+1}(x)}\int_{-1}^1\diff z_s\frac{t(s,z_s)^n}{2}\frac{z_s^m\Omega\big(t(s,z_s)\big)}{x-t(s,z_s)}\,.
\end{align}
Equation~\eqref{eq:Fint} is thus a linear integral equation for $\tilde F(s)$, to be solved for a given 
set of subtraction constants. Discretizing Eq.~\eqref{eq:Fint} yields
\begin{align}
A(s_i)&=\sum_j\bigg(\delta_{ij}-\int_{s_j}^{s_{j+1}}K(s_i,x)\,\diff x\bigg)\tilde{\F}(s_j)\,,
\end{align}
which is solved by matrix inversion; the numerical treatment is relegated to Appendix~\ref{app:Numerics}.

\section{Numerical results and experimental comparison}\label{sec:numerics}

\begin{figure}
\centering
\includegraphics*[width=\linewidth]{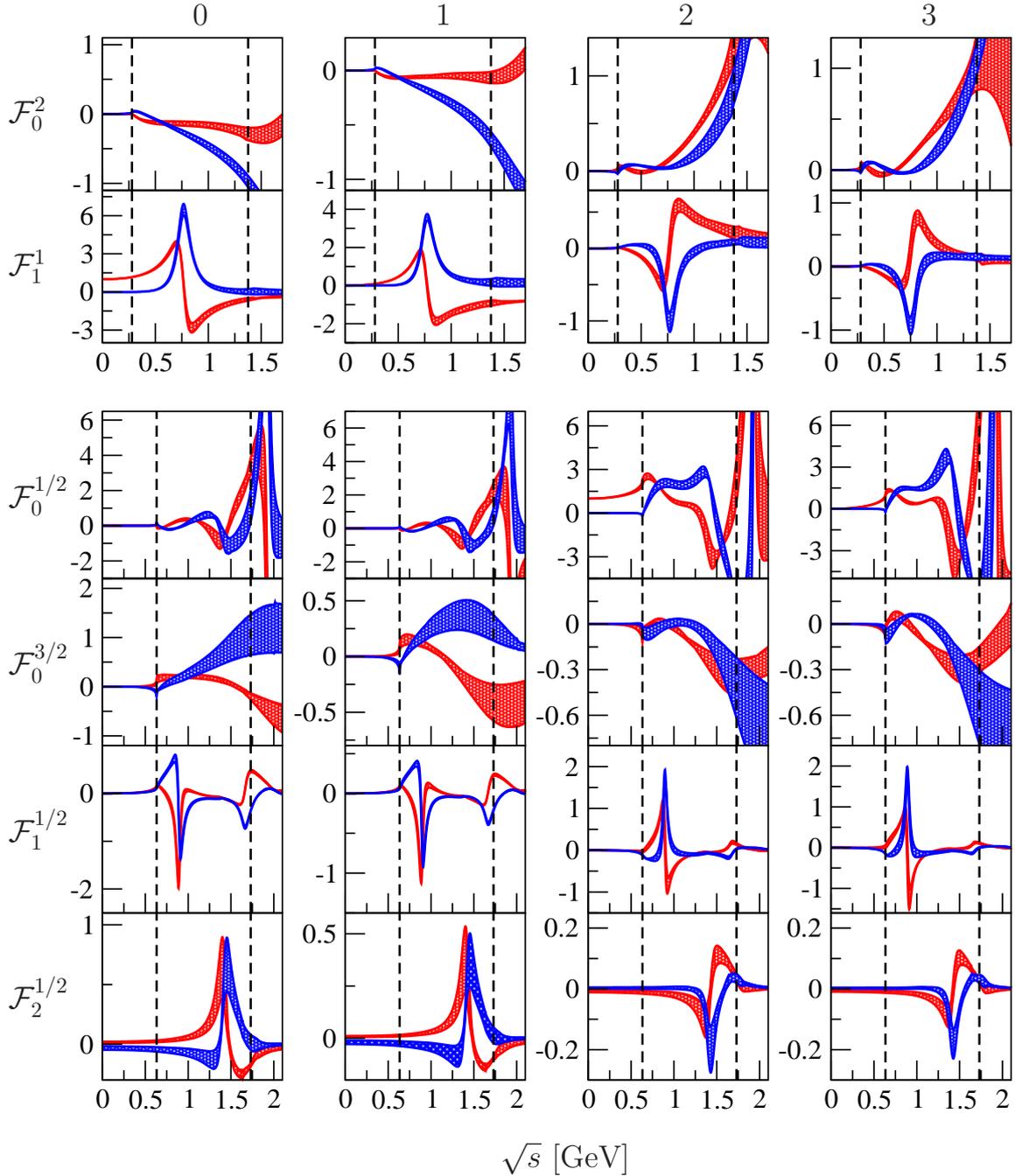}
\caption{Real (red) and imaginary (blue) parts of the single-variable functions $(\F_L^I)_i$ for $i=0,\,\ldots,\,3$.
The vertical dashed lines denote the kinematical limits of the decay region.}
\label{fig:basis1}
\end{figure}
\begin{figure}
\centering
\includegraphics*[width=\linewidth]{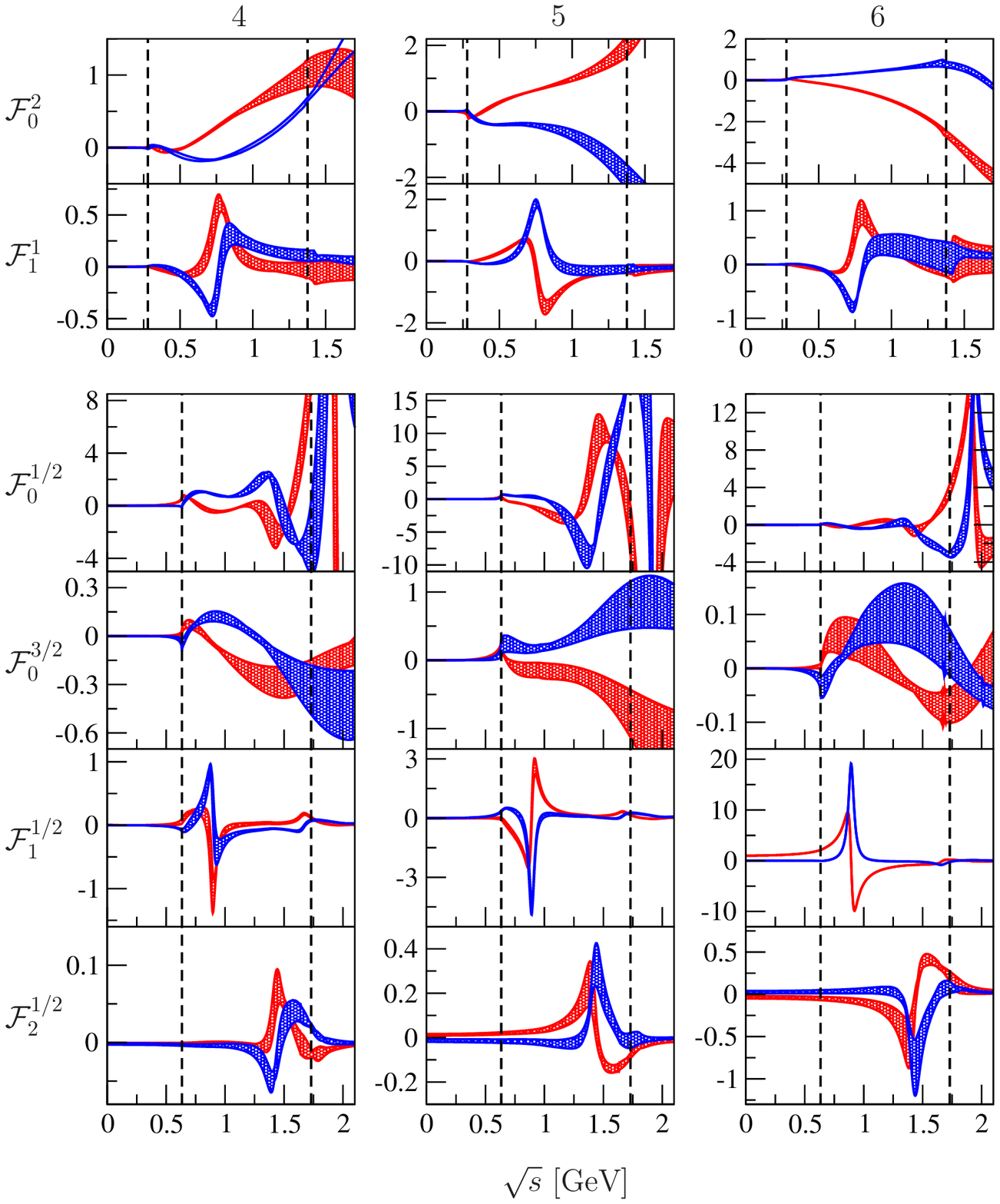}
\caption{Real (red) and imaginary (blue) parts of the single-variable functions $(\F_L^I)_i$ for $i=4,\,\ldots,\,6$.
The vertical dashed lines denote the kinematical limits of the decay region.}
\label{fig:basis2}
\end{figure}
Solving the coupled integral Eqs.~\eqref{eq:fulleq} with the algorithm presented in the previous section, we
obtain the single-variable basis functions $(\F_L^I)_i$ depicted in Figs.~\ref{fig:basis1} and \ref{fig:basis2}.
The vector resonances $K^*(892)$ (in $\F^{1/2}_1$) and $\rho(770)$ (in $\F^1_1$) as well as the $\pi K$ $D$-wave resonance $K_2^*(1430)$ (in $\F^{1/2}_2 $) are clearly visible.
The $\F^{1/2}_0$ basis functions include the effects of the scalar states $K_0^*(800)$ and $K_0^*(1430)$, while
the exotic $\F^2_0$ and $\F^{3/2}_0$ basis functions are free of resonances.

The error bands in  Figs.~\ref{fig:basis1} and \ref{fig:basis2} are determined by a conservative error estimate of the phase shifts:
For the $S$-wave $\pi K$ and $\pi\pi$ phases the error is assumed to rise linearly from zero at threshold to $\pm20^{\circ}$ at 2~GeV. Beyond 2~GeV the error is fixed to $\pm20^{\circ}$.
The $\pi K$ isospin $1/2$ $P$- and $D$-wave phase errors and $\pi\pi$ $P$-wave phase errors are similarly obtained, with the only difference that the linear rise of the error sets in after the $K^*(892)$, $K^*_2(1430)$, and $\rho(770)$ resonances, respectively.
In the $\pi\pi$ $P$-wave case we additionally vary between the phase-shift data from Refs.~\cite{ACGL,Caprini:2011ky,CapriniWIP,Pelaez}.

In the following we compare our theoretical decay amplitude to the experimental $D^+\to K^-\pi^+\pi^+$ Dalitz plot data from the CLEO~\cite{CLEO} and FOCUS~\cite{FOCUS} collaborations. 
Exploiting the symmetry of the process under the interchange of the two pions, we can restrict the comparison to the region $s<t$ by mirroring the remaining half of the Dalitz plot into this region.

\begin{sloppypar}
The experimental events are collected in equidistant bins of size $0.044\,\text{GeV}^2\times 0.044\,\text{GeV}^2$. Bins which overlap with the phase space boundary are discarded, resulting in 493~bins over the considered fit region ($s<t<(\mK+\metap)^2$).
The following event distribution function was used for the fit analogously to the experimental analyses 
\begin{align}\label{eq:PDF}
 \mathcal{P}(s_i,t_i)=\int_{t_i-\delta}^{t_i+\delta} \int_{s_i-\delta}^{s_i+\delta}
\Big[f_{\text{sig}}\N_S|\M_{-++}(s,t,u)|^2\epsilon(s,t)+(1-f_{\text{sig}})\N_B B(s,t) \Big] \diff s\,\diff t\,,
\end{align}
with $(s_i,t_i)$ being the center of the corresponding bin and $2\delta$ the bin width,  $\epsilon(s,t)$ the efficiency parametrization, $B(s,t)$ the background parametrization, $\N_{\text{sig}}$ and $\N_B$ normalization constants such that the background and signal term are normalized to unity,  and the signal fraction $f_{\text{sig}}$.
\end{sloppypar}

We minimize the following $\chi^2$,
\beq
 \chi^2=\sum_{i=0}^{492} \frac{\big[\N\mathcal{P}(s_i,t_i)-(\text{\#events/bin})_i\big]^2}{(\text{\#events/bin})_i}\,,
\eeq
where $\N$ is the number of events, 
the sum runs over the number of bins and the error on the binned data is assumed to be purely statistical.
In addition to the full dispersive representation Eq.~\eqref{eq:fulleq}, we also fit 
a simplified decay amplitude to data, which is given by a sum of
Omn\`es functions multiplied by polynomials: 
\begin{align}\label{eq:Omdecay}
\M_{-++}(s,t,u)&=c'_0 \Omega_0^2(u) -\sqrt{\frac{2}{15}}c'_1\Omega_0^{3/2}(s)+ \frac{1}{\sqrt{3}}\big(c'_2 +c'_3 s+c'_4 s^2+c'_5 s^3\big)\Omega_0^{1/2}(s) \nnnl
&+\frac{c'_6}{\sqrt{3}} \big[s(t-u)-\Delta\big]\Omega^{1/2}_1(s)+\frac{c'_7}{2\sqrt{3}} \Big[3\big(s(t-u)-\Delta\big)^2-\kappa^2(s)\Big]\Omega^{1/2}_2(s)\nnnl
&+(t\leftrightarrow s)\,,
\end{align}
where the $c'_i$ are again complex fit constants.  Equation~\eqref{eq:Omdecay} emulates
a dispersively improved isobar model that neglects any crossed-channel rescattering effects.
The number of polynomial fit constants is chosen to resemble the number of degrees of freedom
in the full dispersive result Eq.~\eqref{eq:fulleq} as far as possible; with certain caveats
that preclude an immediate quantification of three-particle rescattering effects
in the same straightforward way as performed for $\phi\to3\pi$ decays in Ref.~\cite{Niecknig}.
In Eq.~\eqref{eq:fulleq}, two subtraction constants $c_0$ and $c_1$ are contained in the $\pi\pi$ $P$-wave, which only contributes indirectly via the intermediate state $\bar{K}^0\pi^0\pi^+$ to the decay and thus does not show up in the pure Omn\`es amplitude Eq.~\eqref{eq:Omdecay}. In addition, every Omn\`es function 
in Eq.~\eqref{eq:Omdecay} needs at least a normalization constant to adjust the strength of individual amplitudes, while some single-variable amplitudes do not have any subtraction constants. 
Finally, once the $D$-wave is included we have one additional complex fit parameter $c'_7$ in the pure Omn\`es fits. For that reason we consider both Omn\`es and the full dispersive fits without (Omn\`es~1, full~1) and with $D$-wave (Omn\`es~2, full~2).

We have the freedom to fix one subtraction constant, as both the overall normalization and the overall phase are arbitrary and factorized out;
we choose $c_2=c'_2=1$. This leaves $13$ $(15)$ real fit constants for the full / Omn\`es fits.

Following experimental custom, we will employ so-called fit fractions to characterize the relative 
importance of various single-variable functions.
These are defined in the following way
\beq \label{eq:fitfrac}
\FF^I_J=\frac{\int|P_J(x(s,t))\,\F^I_{J}(x(s,t))|^2\, \diff s\,\diff t}{\int |\M_{-++}(s,t,u)|^2\, \diff s\,\diff t}\,, 
\eeq
where the $P_J$ denote the angular prefactors of the corresponding single-variable amplitudes in the total amplitude.
The integration runs over the fitted Dalitz plot region.
In general these fit fractions are not unique due to the freedom of adding an element of the 
invariance group Eq.~\eqref{eq:invgroup};
the projections onto partial-wave amplitudes then will lead to different fit fractions.

\subsection{Comparison to the CLEO data}
\begin{table}
\centering
\renewcommand{\arraystretch}{1.4}
\begin{tabular}{c c c c c c}
\toprule
			       & Full 1				& Full 2 			&
                               & Omn\`es 1  	                & Omn\`es 2 		\\  
\midrule
$|c_0|\times\text{GeV}^2$	       &$2.7\pm0.8$  			&$1.2\pm0.2$			&
$|c'_0|$                         &$0.9\pm0.3$ 	                &$0.9\pm0.7$ 		\\  
$|c_1|\times\text{GeV}^4$      &$3.8\pm1.2$		 	&$2.2\pm0.5$ 			&
$|c'_1|$                       & $3.0\pm1.5$			&$4.0\pm1.3$		\\  
$c_2$ 		       & 1 (fixed) 			& 1 (fixed) 			&
$c'_2$                       & 1 (fixed) 		        & 1 (fixed) 		\\  
$|c_3|\times\text{GeV}^2$      &$2.8\pm0.4$			& $2.2\pm0.1$			&
$|c'_3|\times\text{GeV}^2$     &$1.9\pm0.2$			&$2.0\pm0.2$		\\  
$|c_4|\times\text{GeV}^4$      &$2.0\pm0.5$    			&$1.4\pm0.1$			&
$|c'_4|\times\text{GeV}^4$     &$0.9\pm0.1$ 		        &$1.1\pm0.1$       \\  
$|c_5|\times\text{GeV}^6$      &$0.7\pm0.3$	     	 	&$0.4\pm0.1$		&
$|c'_5|\times\text{GeV}^6$     &$0.13\pm0.3$			&$0.19\pm0.02$		\\ 
$|c_6|\times10^2\text{GeV}^4$  &$4\pm 3$			&$6\pm 2$	&
$|c'_6|\times\text{GeV}^4$     &$0.11\pm0.05$		  	&$0.10\pm0.03$	\\ 
			       &				&				&
$|c'_7|\times10^3\text{GeV}^8$ &---				&$6\pm4$		\\  
arg\,$c_0$		       &$0.1\pm0.2$ 			&$1.1\pm0.3$			&
arg\,$c'_0$		       &$0.2\pm0.8$ 			&$0.4\pm0.4$	\\ 
arg\,$c_1$		       &$0.3\pm0.2$  			&$1.2\pm0.3$			&
arg\,$c'_1$ 		       &$-0.8\pm0.3$			&$-0.4\pm0.2$		\\  
arg\,$c_3$		       &$-0.2\pm0.1$ 			&$0.0\pm0.1$  		&
arg\,$c'_3$ 		       &$0.2\pm0.2$			&$0.3\pm0.2$		\\  
arg\,$c_4$		       &$-0.5\pm0.1$			&$0.0\pm0.1$ 		&
arg\,$c'_4$ 		       &$0.4\pm0.2$			&$0.2\pm0.2$	\\  
arg\,$c_5$		       &$-0.1\pm0.1$			&$0.1\pm0.1$ 			&
arg\,$c'_5$ 		       &$0.2\pm0.4$			&$0.0\pm0.3$		\\  
arg\,$c_6$		       &$-0.3\pm1.2$			&$-0.9\pm0.2$  		&
arg\,$c'_6$ 		       &$0.0\pm0.1$			&$0.0\pm0.3$		\\  
                               &				&				&
arg\,$c'_7$ & --- 	       &$0.4\pm0.3$		\\  
\midrule
$\chi^2/\text{d.o.f.}	$  	&$1.18 \pm 0.03$ 		&$1.10 \pm 0.02$			& &$1.30\pm0.06$	&$1.08\pm0.02$	\\  
\bottomrule
\end{tabular}
\renewcommand{\arraystretch}{1.0}
\caption{{\it Fit to CLEO data:} Numerical fit results for the subtraction constants $c_i$ and $c'_i$ and the corresponding $\chi^2/\text{d.o.f.}$. Four fit scenarios are considered: the full dispersive fit, without $D$-wave (full 1) and with $D$-wave (full 2), 
and the Omn\`es fits of Eq.~\eqref{eq:Omdecay}, without $D$-wave (Omn\`es 1) and with $D$-wave (Omn\`es 2). The errors on the parameters are evaluated by varying the basis functions within their error bands.}
\label{tab:tabFITresult}
\end{table}

\begin{table}
\centering
\renewcommand{\arraystretch}{1.4}
\begin{tabular}{c c c c c c c}
\toprule
Fit  		&$\FF_0^2$			&  $2\times \FF_0^{1/2}$		&  $ 2\times \FF_1^{1/2}$&  $ 2\times \FF_0^{3/2}$ 	&  $ 2 \times \FF_2^{1/2}$ 	\\  
\midrule
Full 1 		&$(37\pm 23) \%$		&  $(190 \pm 60)\%$ 		&  $(11\pm 3)\%$ 	&  $(65\pm 35)\%$ 		&  ---   			\\  
Full 2 		&$(8\pm 3)\%$			&  $(72\pm12)\%$ 		&  $(10\pm 2)\%$ 	&  $(16\pm 3)\%$ 		&  $(0.1\pm0.05)\% $ 		\\ 
Omn\`es 1 	&$(48 \pm 16) \%$		&  $(178 \pm 22)\%$ 		&  $( 7\pm 1)\%$ 	&  $(395 \pm 35) \%$ 		&  ---  				\\  
Omn\`es 2 	& $(9.5\pm 8)\%$		&  $(91\pm 22)\%$ 		&  $(8\pm0.5)\%$ 	&  $(240 \pm 40)\% $ 		& $(0.13\pm0.03)\% $  		\\  
\bottomrule
\end{tabular}
\renewcommand{\arraystretch}{1.0}
\caption{{\it Fit fractions CLEO:} The resulting fit fractions of Eq.~\eqref{eq:fitfrac} for the different fit scenarios; the errors on the parameters are evaluated by varying the basis functions within their error bands.
The fit fractions for the $\pi K$ amplitudes  are multiplied by two to account for the $s\leftrightarrow t$ symmetry.}
\label{tab:tabFitfrac}
\end{table}

\begin{figure}
\centering
\includegraphics*[width=0.6\linewidth]{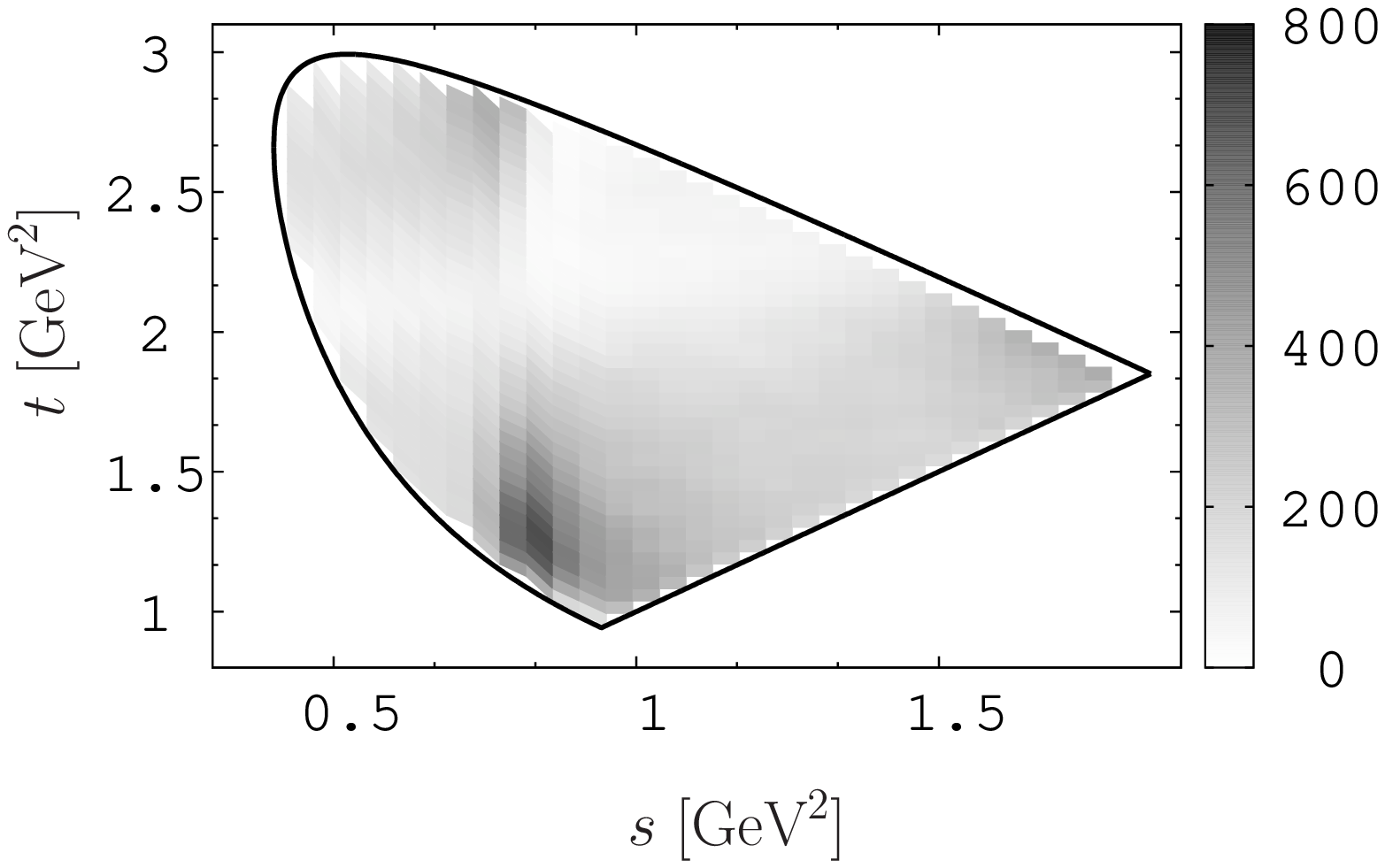}\\
\includegraphics*[width=0.6\linewidth]{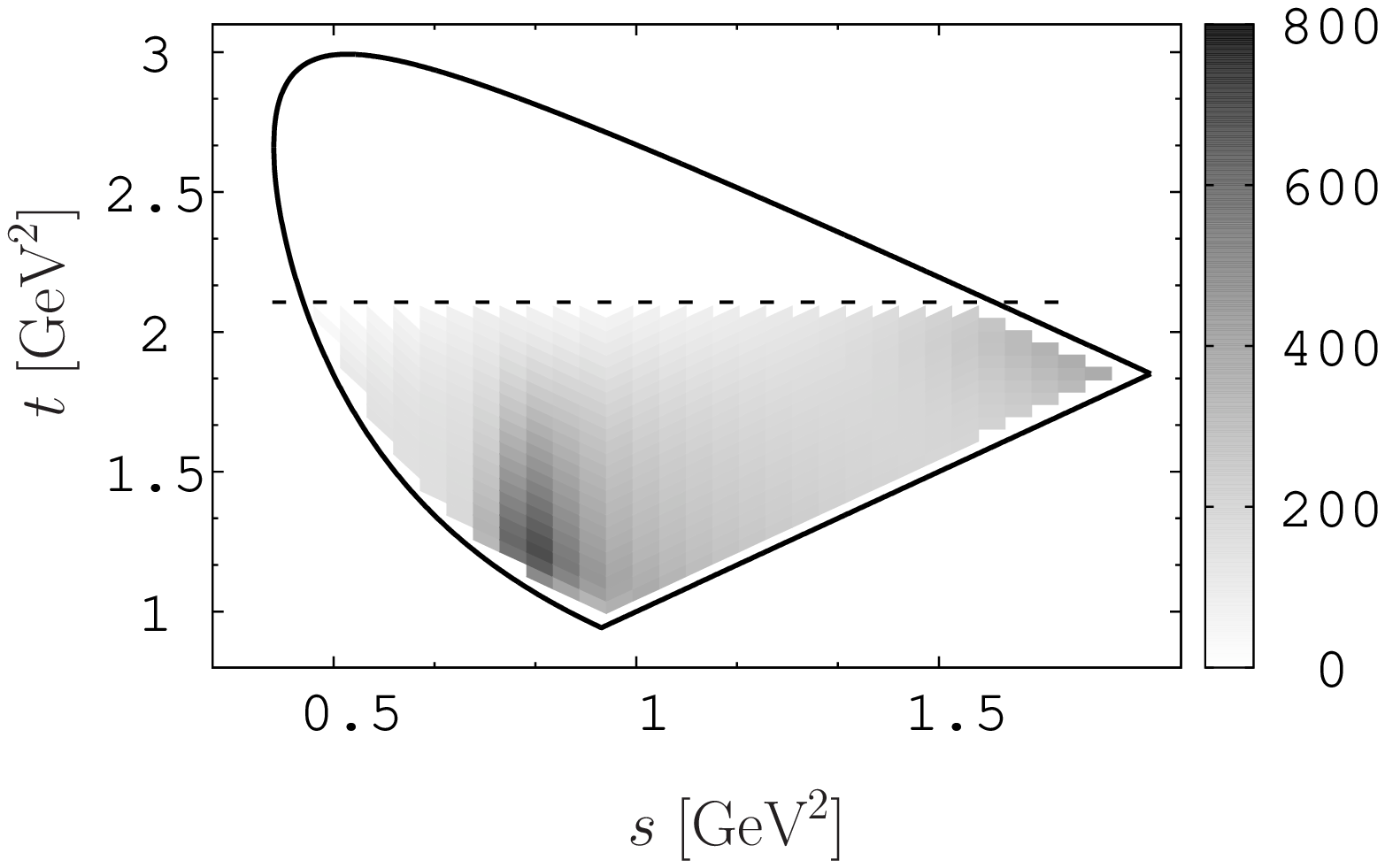}\\[1mm]
\includegraphics*[width=0.71\linewidth]{slicet.eps}
\caption{{\it From top to bottom:} The experimental data from CLEO~\cite{CLEO} depicted in a binned Dalitz plot. 
Below that the theoretical Dalitz plot fitted to the data (fit~2). 
The dashed line denotes the $\eta' K$ threshold. 
The lowest plots show slices through the Dalitz plot. The red and blue curves correspond to the full fits 1 and 2, respectively.}
\label{fig:Dalitz}
\end{figure}

The Dalitz plot measured by the CLEO collaboration~\cite{CLEO} contains 140793~events.
The efficiency and background parametrizations are given explicitly.\footnote{
The threshold factors $T(x)$ used in there read~\cite{Dubrovin:privcomm}
\begin{align*}
 T(x)=\left\{\begin{array}{ll}\sin\big(\pi E_{{\rm th},x}|x-x_{\rm max}|\big), & \text{for } 0<E_{{\rm th},x}|x-x_{\rm max}|<1/2\,, \\
              1, & \text{for } E_{{\rm th},x}|x-x_{\rm max}|\geq 1/2\,.
             \end{array}\right. 
\end{align*}
}
Our fit results are summarized in Table~\ref{tab:tabFITresult}, together with the fit fractions in Table~\ref{tab:tabFitfrac}.
In the full dispersive fits (full fits~1/2), the resulting values for the  subtraction constants in Table~\ref{tab:tabFITresult} have similar order of magnitude with the exception $c_6$, which is rather small.
This can be understood by the large $\F_1^{1/2}$ single-variable amplitude in this particular basis function (see Fig.~\ref{fig:basis2}).
Furthermore the phases of the $\F_1^1$ subtraction constants ($c_0$, $c_1$) nearly agree modulo $\pi$. The same holds for the  $\F_0^{1/2}$ subtraction constants ($c_2$ to $c_5$) especially for the full fit 2.
This suggests that with overall phases factorized, the subtraction constants for the $\F_1^1$ and likewise the $\F_0^{1/2}$ amplitude are almost real.
The differences of the single-variable amplitude phases to the elastic phase shifts depicted in Fig.~\ref{fig:PW} are thus predominantly due to the dispersion integrals, i.e.\ the crossed-channel rescattering effects.

Including the $D$-wave improves the $\chi^2/\text{d.o.f.}$ slightly from $1.18\pm0.03$ to $1.10\pm 0.02$. 
Note that in the full dispersive representation, no additional 
fit constants are introduced when the $D$-wave is added.
The inclusion of the $D$-wave does not change the phases of most subtraction constants
beyond their uncertainties, with the exception of $c_6$; the magnitudes, in contrast, 
change significantly for almost all subtractions.
Considering the fit fractions in Table~\ref{tab:tabFitfrac}, we observe that the inclusion of the $D$-wave in the full fit 2 reduces the highly destructive interference between the two $S$-wave amplitudes in the $\pi K$ channel.
We wish to point out that also in Ref.~\cite{FOCUS}, a large cancellation between the isospin 1/2 and isospin 3/2 $S$-wave components of $-164\%$ is seen, with individual fit 
fractions of $(207\pm24)\%$ and $(40\pm9)\%$, respectively, which show a comparable behavior to our full fit~1. 
Although the fit fraction of the $D$-wave itself is very small, it thus has a rather large impact on the $S$- and $P$-waves. A similar phenomenon is seen in Ref.~\cite{CLEO} 
where the fit quality deteriorates considerably when removing the small $D$-wave. Although we do not fit the whole Dalitz plot, the fit fractions for the resonant single-variable amplitudes 
for $\F^{1/2}_0$, $\F^{1/2}_1$ and $\F^{1/2}_2$ agree well with the results from Refs.~\cite{CLEO,FOCUS}.
The $\F_0^2$ fit fraction corresponds to the isospin 2 $\pi\pi$ $S$-wave component of $\FF\approx(9.8\dots 15.5)\%$ found in Ref.~\cite{CLEO} within different fit models,
and together with the fit fraction of $\F_0^{3/2}$ agrees with the nonresonant contribution found in Ref.~\cite{FOCUS} of $\FF\approx (29.7\pm4.5)\%$.

Although the Omn\`es fits (Omn\`es 1, 2) yield overall similar $\chi^2$ results, the strengths of the individual amplitudes shown in Table~\ref{tab:tabFitfrac} are highly implausible and probably sufficient to
reject this model. 
In particular the contribution of the nonresonant isospin 3/2 $\pi K$ $S$-wave is vastly beyond all reasonable
expectations, and cannot be justified.  In contrast to the full fit, this situation is not ameliorated significantly
by including the $D$-wave.
We conclude that crossed-channel rescattering effects are essential to obtain sensible fit fractions.

The resulting Dalitz plot as well as a one-dimensional representation in terms of
slices through it are displayed in Fig.~\ref{fig:Dalitz}.
The bin numbering for the latter is organized in terms of $t$-slices for constant $s$,
subsequently glued together with the next slice of higher $s$.
We evaluate the event distribution function Eq.~\eqref{eq:PDF} over each bin and compare to experimental data.
The rather small error band on the fit results suggests that the uncertainty in the basis functions is largely 
compensated by interference effects between the different single-variable amplitudes, as well as
by corresponding variations in the fitted subtraction constants.

\subsection{Comparison to the FOCUS data}
\begin{table}
\centering
\renewcommand{\arraystretch}{1.4}
\begin{tabular}{c c c c c c}
\toprule
Fit constant 			&Full 1				& Full 2 			&
                                &Omn\`es 1  	                & Omn\`es 2 		\\ 
\midrule
$|c_0|\times\text{GeV}^2$	        &$3.0\pm0.8$  	 		&$0.6\pm0.3$			&
$|c'_0|$			        &$0.4\pm0.2$ 			&$0.6\pm0.3$		\\  
$|c_1|\times\text{GeV}^4$ 	&$3\pm1$			& $0.9\pm0.3$			&
$|c'_1|$ 			&$1.9\pm0.8$	 	        &$2.2\pm0.5$		\\  
$c_2$ 	                & 1 (fixed) 	 		& 1 (fixed) 			&
$c'_2$ 		        & 1 (fixed) 			& 1 (fixed) 		\\  
$|c_3|\times\text{GeV}^2$ 	&$2.8\pm0.8$ 			&$1.9\pm0.1$ 			&
$|c'_3|\times\text{GeV}^2$ 	&$1.7\pm0.2$	                &$1.8\pm0.2$		\\  
$|c_4|\times\text{GeV}^4$	&$2.5\pm0.6$		  	&$1.1\pm0.1$			&
$|c'_4|\times\text{GeV}^4$ 	&$0.9\pm0.2$ 	     		&$1.0\pm0.2$		\\  
$|c_5|\times\text{GeV}^6$ 	&$0.4\pm0.2$ 			& $0.3\pm0.1$ 		&
$|c'_5|\times\text{GeV}^6$	&$0.1\pm0.2$ 	     		&$0.3\pm0.1$		\\  
$|c_6|\times\text{GeV}^4$ 	&$0.2\pm0.1$ 			& $0.0\pm0.1$          	&
$|c'_6|\times\text{GeV}^4$ 	&$0.1\pm0.4$		        &$0.1\pm0.1$		\\  
				&   				& 				&
$|c'_7|\times 10^3\text{GeV}^8$ &--- 				&$7\pm4$ 			\\  
arg\,$c_0$		  	&$0.5\pm0.3$ 			&$0.9\pm0.3$			&
arg\,$c'_0$		  	&$0.7\pm0.5$ 			&$0\pm1$		\\  
arg\,$c_1$		  	&$0.6\pm0.4$ 			&$1.1\pm0.2$			&
arg\,$c'_1$		  	&$-1.1\pm0.4$			&$0.2\pm0.3$		\\  
arg\,$c_3$		  	&$0.0\pm0.2$ 			&$0.0\pm0.1$			&
arg\,$c'_3$		  	&$0.4\pm0.2$			&$0.2\pm0.2$		\\ 
arg\,$c_4$		 	&$-0.2\pm0.3$ 			&$0.0\pm0.1$			&
arg\,$c'_4$		  	&$0.6\pm0.2$			&$0.2\pm0.3$	\\  
arg\,$c_5$		  	&$0.2\pm0.3$			&$0.0\pm0.1$ 			&
arg\,$c'_5$		  	&$0.8\pm0.2$			&$0.2\pm0.3$		\\  
arg\,$c_6$		  	&$-0.6\pm0.7$ 			&$-1.0\pm0.4$  			&
arg\,$c'_6$		  	&$-0.7\pm0.3$			&$-0.9\pm0.3$		\\  
        		  	&				& 				&
arg\,$c'_7$		  	& --- 		&$-1.1\pm0.5$		\\  
\midrule
$\chi^2/\text{d.o.f.}	$  	&$1.20 \pm 0.01$ 		&$1.21\pm0.02$			&
& $1.25\pm0.02$	&$1.17\pm0.01$	\\  
\bottomrule
\end{tabular}
\renewcommand{\arraystretch}{1.0}
\caption{{\it Fit to FOCUS data:} Numerical fit results for the subtraction constants $c_i$ and $c'_i$ and the corresponding $\chi^2/\text{d.o.f.}$. The same four fit scenarios as in Table~\ref{tab:tabFITresult} are considered. The errors on the parameters are evaluated by varying the basis functions within their error bands.}
\label{tab:tabFITresult_FOCUS}
\end{table}

\begin{table}
\centering
\renewcommand{\arraystretch}{1.4}
\begin{tabular}{c c c c c c c}
\toprule
Fit  		&$\FF_0^2$			&  $2\times \FF_0^{1/2}$		&  $ 2\times \FF_1^{1/2}$&  $ 2\times \FF_0^{3/2}$ 	&  $ 2 \times \FF_2^{1/2}$ 	\\  
\midrule
Full 1 		&$(12\pm 4) \%$			&  $(59\pm 25)\%$ 		&  $(7.5\pm 2.5)\%$ 	&  $(39\pm 27)\%$ 		&  ---  			\\  
Full 2 		&$(5\pm 3)\%$			&  $(67\pm10)\%$ 		&  $(12\pm 1)\%$ 	&  $(8\pm 6)\%$ 		&  $(0.17\pm0.07)\% $ 		\\  
Omn\`es 1 	&$(33\pm 17 ) \%$		&  $(91\pm 37 )\%$ 		&  $(9\pm 1 )\%$ 	&  $(215\pm 135 ) \%$ 		&  --- 				\\ 
Omn\`es 2 	& $(89\pm 42 )\%$		&  $(20\pm12 )\%$ 		&  $(11\pm 1)\%$ 	&  $(180 \pm 60)\% $ 		& $(0.4\pm 0.05)\% $  		\\  
\bottomrule
\end{tabular}
\renewcommand{\arraystretch}{1.0}
\caption{{\it Fit fractions FOCUS:} The resulting fit fractions of Eq.~\eqref{eq:fitfrac} for the different fit scenarios; the errors on the parameters are evaluated by varying the basis functions within their error bands.
The fit fractions for the $\pi K$ amplitudes are multiplied by two to account for the $s\leftrightarrow t$ symmetry.}
\label{tab:tabFitfrac_FOCUS}
\end{table}

\begin{figure}
\centering
\includegraphics*[width=0.6\linewidth]{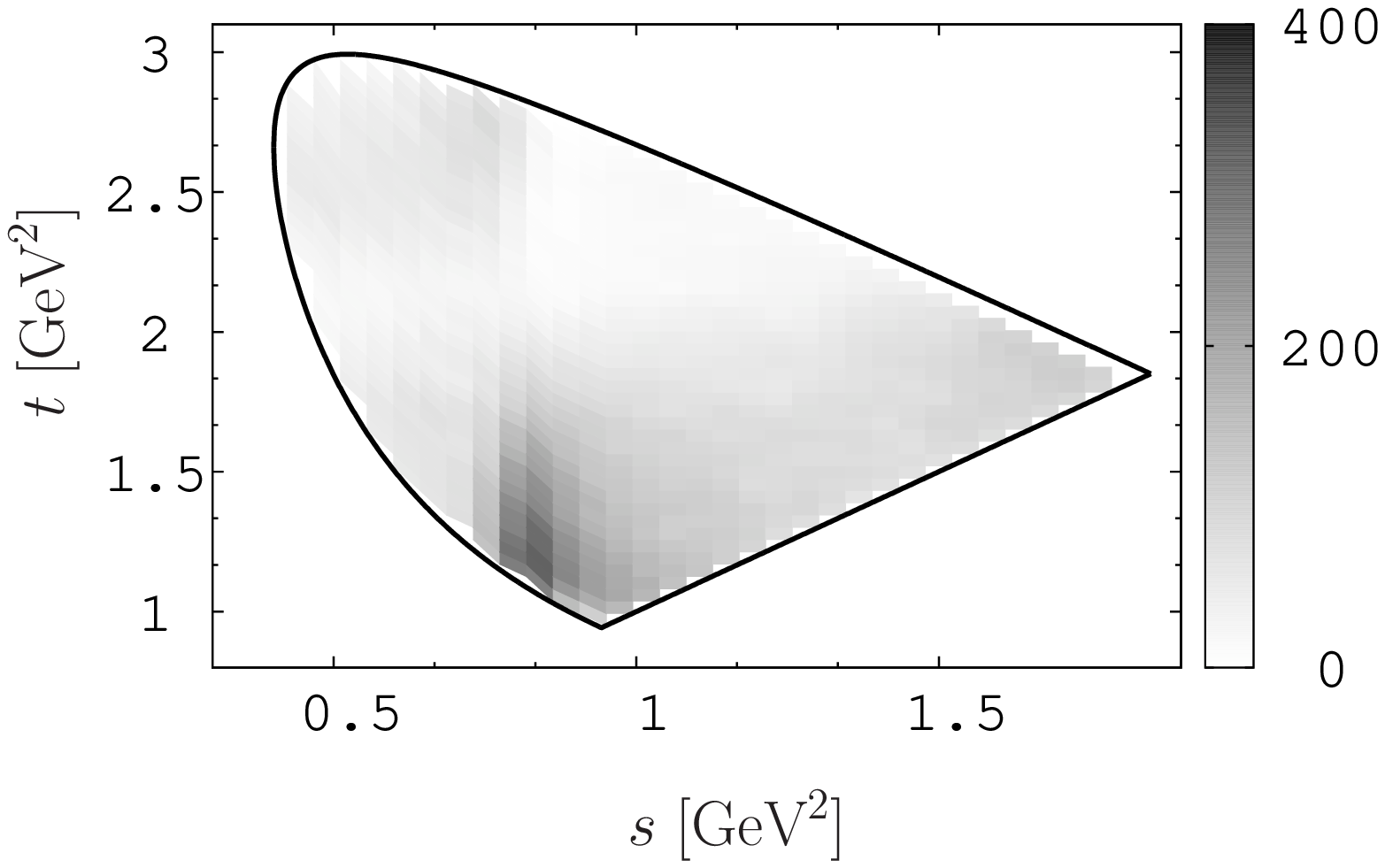}\\
\includegraphics*[width=0.6\linewidth]{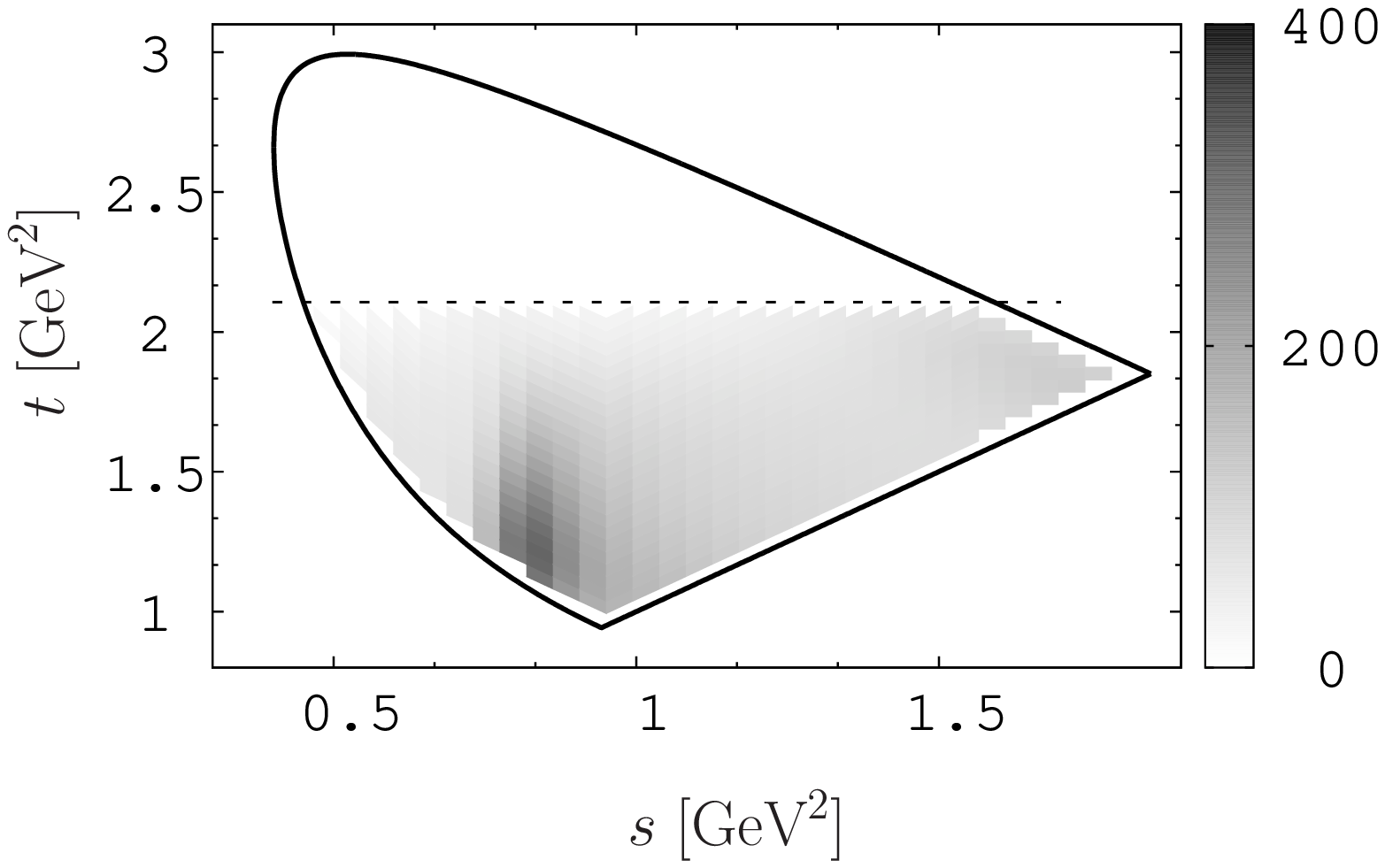}\\[1mm]
\includegraphics*[width=0.71\linewidth]{slicet_FOCUS.eps}
\caption{{\it From top to bottom:} The experimental data from FOCUS~\cite{FOCUS} depicted in a binned Dalitz plot. 
Below that the theoretical Dalitz plot fitted to the data (fit~2). 
The dashed line denotes the $\eta' K$ threshold. 
The lowest plots show slices through the Dalitz plot. The red and blue curves correspond to the full fits 1 and 2, respectively.}
\label{fig:Dalitz_FOCUS}
\end{figure}
The FOCUS Dalitz plot data~\cite{FOCUS} includes $52460~\pm~245$ signal and $1897~\pm~39$ background events. With the resulting signal fraction of $\sim 96.5\%$ we perform the full and Omn\`es fits as above.
Table~\ref{tab:tabFITresult_FOCUS}  summarizes the fit results together with the fit fractions in Table~\ref{tab:tabFitfrac_FOCUS}. 
The overall picture is very similar to the CLEO fit results with a slightly bigger $\chi^2/\text{d.o.f.}\approx 1.2$.  
The Omn\`es fits again result in nonphysical fit fractions (see Table~\ref{tab:tabFitfrac_FOCUS}), and from here on we will only compare the full fits of both experimental data sets.
Starting with the fit without $D$-wave (full fit 1) we observe similar moduli of the subtraction constants compared to the CLEO results, however the phases do differ.
The fit does not show the large destructive interference effects between the isospin 1/2 and isospin 3/2 $S$-wave that we find in the CLEO fit.

No improvement in the $\chi^2/\text{d.o.f.}$ is observed when we include the $D$-wave (full fit 2). However the contribution from the nonresonant amplitudes, the isospin 2 and isospin 3/2 $S$-waves, are reduced (see Table~\ref{tab:tabFitfrac_FOCUS}).
The fit fractions of the full fit 2 differ slightly from the CLEO fits; in particular the nonresonant $S$-waves contribute less in the FOCUS data.

In the full fit~2 the phases of the $\F_1^1$ subtraction constants persist to nearly agree modulo $\pi$; 
the same holds for $\F_0^{1/2}$ subtraction constants.
It is reassuring that the overall picture of the phases of various subtraction constants is consistent in the 
full fit~2 results for both CLEO and FOCUS.

\begin{figure}
\centering
\includegraphics[width=\linewidth]{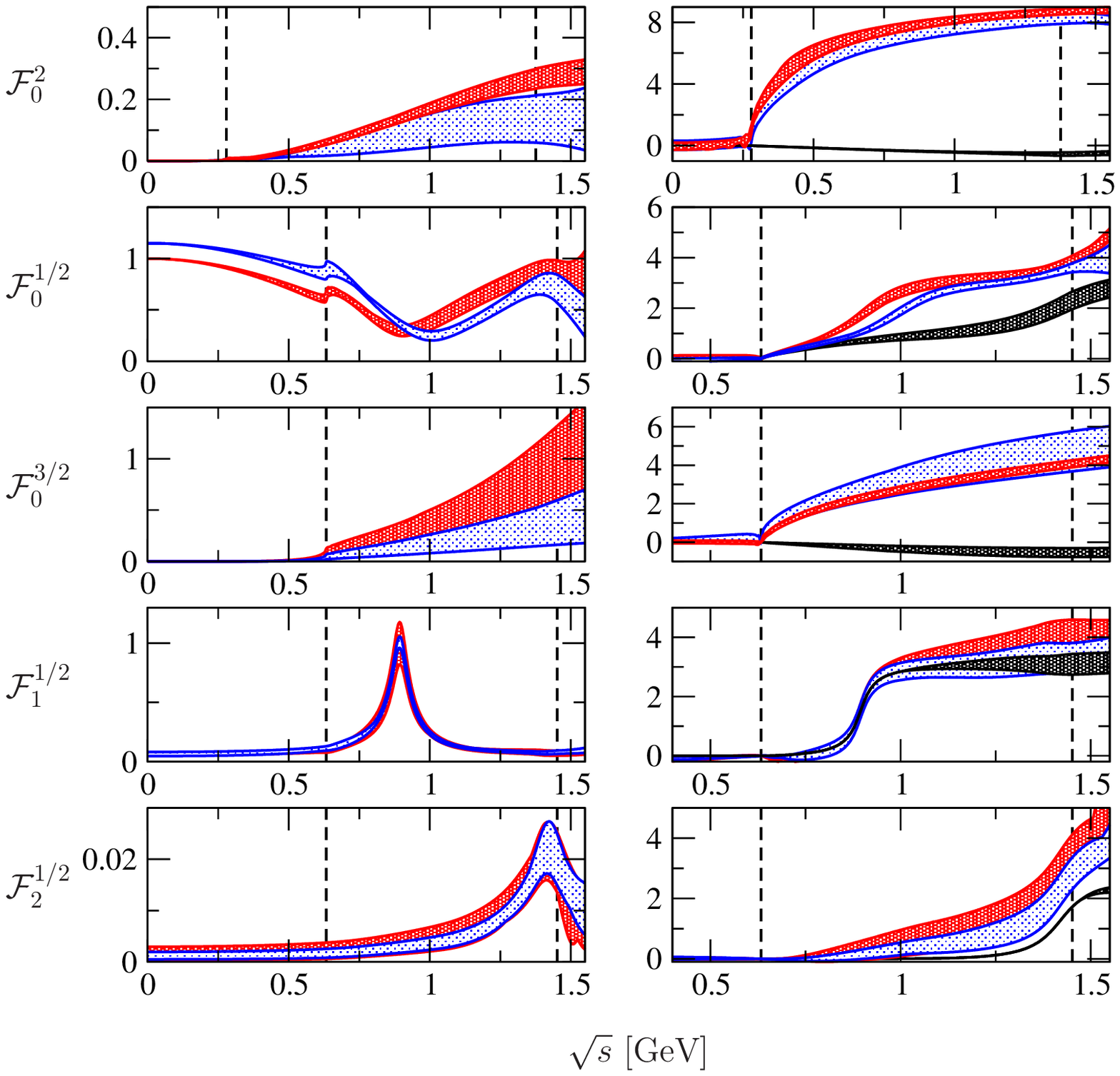}
\caption{{\it Left column:} Absolute values of the single-variable amplitude in arbitrary units of full fit~2 (CLEO in red, FOCUS in blue). The overall normalization is chosen such that the absolute values in the $K^*(892)$ peak agree. 
{\it Right column:} Phases of the single-variable amplitudes (CLEO: red, FOCUS: blue) and input scattering phases (black) in radiant. The phases are fixed to zero at the two-particle ($\pi\pi$, $\pi K$) thresholds.
The dotted lines visualize the fitted area; for the $\pi K$ amplitudes from threshold to the $\eta' K$ threshold  and the full phase space for the $\pi\pi$ amplitudes.}
\label{fig:PW}
\end{figure}

In Fig.~\ref{fig:PW}, we compare moduli and phases of the resulting single-variable amplitudes as fitted to the two
data sets; the phases are also compared to the input phase shifts used in the Omn\`es functions.
The resulting phase motions largely agree in the two analyses within uncertainties, with the possible exception of 
some deviations in $\F_0^{1/2}$ in the region of the $K_0^*(800)$ resonance, where the phase extracted from the CLEO fit rises
more quickly.  There are significant deviations from the input phase shifts throughout: there is no naive realization
of Watson's theorem in the presence of three-body rescattering effects, see e.g.\ recent discussions in 
Refs.~\cite{JLab:methods,Schneider:2012ez}. 
This is also the explanation for the observed discrepancy of the $\pi K$ $I=1/2$ $S$-wave phase as extracted from these 
decays by the E791~\cite{Aitala:2005yh} and FOCUS~\cite{Link:2009ng} collaborations, compared to the 
scattering phase-shift analyses~\cite{Aston:1987ir}:
while the phase shift rises to about $67^\circ$--$97^\circ$ at $\sqrt{s}=1.3\,\text{GeV}$~\cite{piK}, the 
experimental analyses of $D$-decay data
suggest an increase in the phase from threshold by about $133^\circ$--$164^\circ$ (read off via Ref.~\cite{Magalhaes:2011sh}).
Figure~\ref{fig:PW} shows that in the dispersive formalism, the phase at $1.3\,\text{GeV}$ is about $182^\circ$--$198^\circ$ (CLEO) or
$170^\circ$--$183^\circ$ (FOCUS)---even larger than found in Refs.~\cite{Aitala:2005yh,Link:2009ng}.  
We emphasize that these results are based on a formalism that uses the scattering phase shifts~\cite{piK} as input: the deviations
in the decay amplitude $S$-wave are due to complex phases induced by three-body rescattering effects.

In general, the corrections compared to input phase shifts
are smallest for narrow resonances, in particular in the $I=1/2$ $\pi K$ $P$- and $D$-waves.
The largest phase differences are observed in the nonresonant amplitudes, where the  phases of $\F_0^2$ and $\F_0^{3/2}$ show a $2\pi$ rise due to zeros in imaginary or real parts close to threshold in individual basis functions.
Note how these seemingly drastic differences are accompanied by very small absolute magnitudes
of the amplitudes in question: in view of the aim to control the phase behavior of the complete, combined decay amplitude
accurately, these specific deviations are still rather small.

Turning to the moduli of the single-variable amplitudes, the relative strength of the $K^*(892)$ resonance (in $\F^{1/2}_1$) compared with the $K^*_2(1430)$ (in $\F^{1/2}_2$) agrees between CLEO and FOCUS fits.
However the dip in the $\F^{1/2}_0$ amplitude is shifted to higher energies in the FOCUS fit and slightly more pronounced.
The moduli of the nonresonant amplitudes $\F^{3/2}_0$ and $\F^2_0$ turn out to be smaller in the FOCUS fit, which is also underlined by the fit fractions (compare Tables~\ref{tab:tabFitfrac} and \ref{tab:tabFitfrac_FOCUS}).

\subsection{Comparison to other approaches}

So far the only theoretical approach known to us that includes all relevant partial waves, three-particle rescattering effects, and the isospin coupled intermediate state $\ol{K}^0\pi^0\pi^+$ is Ref.~\cite{Nakamura:2015qga}.
The treatment is based on a unitary coupled-channel framework. The two-particle rescattering contributions are fixed by the $\pi K$ and $\pi\pi$ scattering data, phases, and moduli. Three-body rescattering effects are generated by solving a Faddeev equation.
In addition to the three-body rescattering a three-body potential, based on  hidden local symmetry, is introduced modeling vector meson exchanges.
The author studies the influence of individual rescattering contributions by considering different fit scenarios;
crossed-channel rescattering effects and three-body potential turned off (isobar fit), three-body potential turned off (Z fit), and the full fit.
An additional  contact term breaking unitarity is allowed for, which
in the full fit turns out to be negligible.
The decay amplitude depends on 27 to 39 degrees of freedom depending on the considered fit model, which is more than twice  the number of parameters included in our full fit.

To compare the fit fractions obtained in Ref.~\cite{Nakamura:2015qga}, we note that the isobar fit theoretically compares closest to our Omn\`es fits, while the Z fit does to our full fits.
However the isobar fit has a large contribution from the unitarity-breaking contact term (considered as a ``background'' contribution) of $17.7\%$, such that a direct comparison is not sensible.
Concerning the full and the Z fit, a large destructive interference between the isospin 1/2 and isospin 3/2 $S$-waves is seen, similar to our CLEO fit~1 configuration.
The isospin 1/2 $P$-waves are of similar size, $\sim 15\%$ compared to our $10-14\%$, but the $\pi\pi$ $S$-wave contribution is smaller ($1.8-3.8 \%$) than our contributions in either full fit~1 or CLEO full fit 2.
It agrees only with the FOCUS full fit~2.
--- Concerning this comparison, we should stress once more that in contrast to Ref.~\cite{Nakamura:2015qga},
we do not fit the full Dalitz plot.

Unfortunately the improvement due to crossed-channel rescattering cannot be quantified in a simple way in Ref.~\cite{Nakamura:2015qga} either.
The improvement going from the isobar to the Z and then further to the full model can also be due to the introduction of further degrees of freedom;
as discussed above, we encounter a similar problem in our analysis. 
However the background term, which gives an indication for missing physics, reduces dramatically once the crossed-channel rescattering effects and the coupled intermediate state $\ol{K}^0\pi^0\pi^+$ are included.
This is a similar conclusion as drawn from the dispersive analysis of $\phi\to3\pi$ Dalitz plots~\cite{Niecknig}, 
which rendered phenomenological contact terms~\cite{Aloisio:2003ur,Akhmetshin:2006sc} superfluous.

\section{Conclusion}\label{sec:conclusion}

In this paper we have analyzed the $D^+\to K^-\pi^+\pi^+$ decay with a dispersive framework based on the Khuri--Treiman formalism that satisfies analyticity, unitarity, crossing symmetry, and includes crossed-channel rescattering among the three final-state particles.

The theoretical decay amplitude depends on seven complex subtraction constants, one of which can be absorbed
into overall phase and normalization of the amplitude.  The remaining parameters 
are fitted to the experimental Dalitz plot data from the CLEO~\cite{CLEO} and FOCUS~\cite{FOCUS} collaborations,
restricting the kinematic region to below the $\eta' K$ threshold, where the elastic approximation is assumed to work well.    
We have considered different fit scenarios with (full) and without crossed-channel rescattering effects (Omn\`es), as well as with and without the $\pi K$ isospin $1/2$ $D$-wave.
Although the Omn\`es fits give reasonable $\chi^2/\text{d.o.f.}$, we obtain large destructive interferences between single-variable amplitudes, which manifest themselves in unphysical fit fractions.
The full fits result in good $\chi^2/\text{d.o.f.}$ around $1.1$ for the CLEO data ($1.2$ for the FOCUS data), with sensible fit fractions throughout.
Including the $\pi K$ isospin $1/2$ $D$-wave does not significantly improve the $\chi^2/\text{d.o.f.}$, however the fit fractions of the nonresonant waves are reduced, giving small interference effects between the single-variable
amplitudes.
We have shown that we can describe the $D^+\to K^-\pi^+\pi^+$ Dalitz plot data in the region where we deem elastic unitarity to hold approximately, solely relying on $\pi\pi$ and $\pi K$ scattering phase shift input and exploiting the constraints
of dispersion theory.

Three-body rescattering effects suspends any strict relation between the phase of the decay partial waves and scattering phase shifts:
we have shown that the significantly stronger rise of the $\pi K$ $S$-wave phases, as observed in analyses of these $D$-meson 
decays~\cite{Aitala:2005yh,Link:2009ng} in comparison to phase shift data, can be understood at least qualitatively
in the framework of Khuri--Treiman equations.

We have simultaneously constructed the formalism for the decay $D^+\to \bar{K}^0\pi^0\pi^+$, which is directly related
to $D^+\to K^-\pi^+\pi^+$ by charge exchange and can be constructed from different linear combinations of the same
(isospin) amplitudes.  This second decay channel has recently been measured by the BESIII 
collaboration~\cite{Ablikim:2014cea}.  
A simultaneous analysis of both Dalitz plots will further exploit the predictive power of the dispersive formalism;
due to the direct contribution of the $\pi\pi$ $P$-wave in the $\pi^0\pi^+$ (as opposed to the $\pi^+\pi^+$) final 
state, we expect to find stronger constraints on the subtraction constants featuring directly in the corresponding
amplitude.
The pertinent investigation is in progress~\cite{inprogress}.

\acknowledgments
We are grateful to the CLEO and FOCUS collaborations for providing us with 
the Dalitz plot data of Refs.~\cite{CLEO,FOCUS}, 
and to Mikhail Dubrovin as well as Alberto Correa dos Reis for extremely helpful e-mail communication
concerning details of these analyses.
We would like to thank Christoph Hanhart and Peter Stoffer for useful discussions.
Financial support by DFG and NSFC through funds provided to the Sino--German CRC~110 
``Symmetries and the Emergence of Structure in QCD''
is gratefully acknowledged.

\appendix

\section{Inhomogeneities}\label{app:inhomo}

In this appendix the inhomogeneities are calculated from Eq.~\eqref{eq:inhomodet}.
To demonstrate the procedure we will perform the calculation explicitly in the case of  $f_L^{I=1/2}(s)$,
\beq
 f_L^{1/2}(s)=\sqrt{3}\M_{-++}^{1/2,L}(s)=\kappa^L(s)\big(\F^{1/2}_L(s)+\hat{\F}^{1/2}_L(s)\big)\,. 
\eeq
We start with the projection of the decay amplitude $\M_{-++}$, Eq.~\eqref{eq:Fullamp}, onto isospin eigenstates in the $s$-channel.
We introduce the following crossing matrices:
\beq
 M_s^I \equiv\sum_{I'}X_{st}^{II'}M_t^{I'}\,,\qquad
 M_t^I \equiv\sum_{I'}X_{tu}^{II'}M_u^{I'}\, 
\eeq
and so on,
where $M_x^I$ is the isospin $I$ eigenstate in the $x$-channel and $X_{xy}^{II'}$ the crossing matrix for the transition from channel $y$ to $x$, where $I$ and $I'$ are the matrix component indices.
We obtain the following explicit forms:
\beq
 X_{st}=\frac{1}{3}\left(\begin{array}{cc} 2 & -\sqrt{10} \\  -\sqrt{\frac{5}{2}} & -2 \end{array}\right)=X_{ts}\, ,
\qquad 
X_{us}=\frac{1}{3}\left(\begin{array}{cc} 1  & \sqrt{10} \\  \sqrt{3} & -\sqrt{\frac{6}{5}} \end{array}\right)\,.
\eeq
The $t$-channel and $u$-channel single-variable amplitudes can be split, with the aid of the crossing matrices, into  $I_s=1/2$ and  $I_s=3/2$ contributions,
\begin{align}
\frac{\F^{1/2}_L(t)}{\sqrt{3}}-\sqrt{\frac{2}{15}}\F^{3/2}_L(t)
&=\underbrace{\frac{2}{3\sqrt{3}}\bigg(\F^{1/2}_L(t)-\sqrt{\frac{5}{2}}\F^{3/2}_L(t)\bigg)}_{I_s=1/2}
\underbrace{+\frac{1}{3\sqrt{30}}\big(\sqrt{10}\F^{1/2}_L(t)+4\F^{3/2}_L(t)\big)}_{I_s=3/2},\nnnl
\F^2_0(u)&=\underbrace{\frac{1}{6}\big(\sqrt{3}(t-s)\F^1_1(u)+5\F^2_0(u)\big)}_{I_s=1/2}\underbrace{-\frac{1}{6}\big(\sqrt{3}(t-s)\F^1_1(u)-\F^2_0(u)\big)}_{I_s=3/2}\,,
\end{align}
with $L\in\{0,1\}$. Retaining the $I=1/2$ pieces only, we have 
\begin{align}
\M^{I_s=1/2}_{-++}(s,t,u) &=
\frac{1}{\sqrt{3}}\F^{1/2}_0(s)
+\frac{2}{3\sqrt{3}}\Big(\F^{1/2}_0(t)-\sqrt{\frac{5}{2}}\F^{3/2}_0(t)\Big)
+\frac{1}{\sqrt{3}}\big[s(t-u)-\Delta\big]\F^{1/2}_1(s)\nnnl
&+\frac{2}{3\sqrt{3}}\big[t(s-u)-\Delta\big]\Big(\F^{1/2}_1(t)-\sqrt{\frac{5}{2}}\F^{3/2}_1(t)\Big)\nnnl
&+\frac{1}{6}\big(\sqrt{3}(t-s)\F^1_1(u)+5\F_0^2(u)\big)\,.
\end{align}
Since there is no isospin~1 component in the $u$-channel amplitudes of $\M_{-++}$, the projections onto this specific component  yield zero and
therefore provide an additional cross-check. Similarly no $I_s=1/2$ component should appear in $\M_{\ol{0}0+}$.
We are left with the angular momentum projection. For a compact notation we define the angular average integration by 
\beq
\la z^n\M \ra_{x_y}(y)\equiv\frac{1}{2}\int_{-1}^1 \diff z_y\, z_y^n \M(x(y,z_y))\,.
\eeq
We immediately obtain
\beq
\lla z^n f\rra_{u_s}=\lla z^n f\rra_{u_t},~\lla z^n f\rra_{t_s}=\lla z^n f\rra_{s_t} , ~\text{and } \lla z^n f\rra_{t_u}=(-1)^n\lla z^n f\rra_{s_u} \,.
\eeq
The angular average integration is straightforwardly performed in the scattering region. The continuation to the decay region, where the naive integration would cross the right-hand cut, has been discussed extensively before~\cite{Bronzan,Niecknig}.
We now perform the partial-wave projection
\beq
\M^{I_s=1/2,L}_{-++}(s,t,u)\equiv\frac{2L+1}{2}\int_{-1}^1 \diff z_s P_L(z_s)\M^{I_s=1/2}_{-++}\big(s,t(s,z_s),u(s,z_s)\big) \,,
\eeq
with the Legendre polynomials $P_L(z_s)$.
For the $S$-wave we obtain 
\begin{align}
\sqrt{3}\M_{-++}^{1/2,0}(s)&=\frac{5\sqrt{3}}{6}\lla\F^2_0\rra_{u_s}
+\frac{1}{2}\lla \big(A_s z+D_s\big)\,\F^1_1\rra_{u_s}+\F^{1/2}_0(s)+
\frac{1}{3}\bigg[\lla 2\F^{1/2}_0-\sqrt{10}\F^{3/2}_0\rra_{t_s}\nnnl
&+\lla \big(A_s^2 z^2+B_s z+ C_s\big) \big(2\F^{1/2}_1-\sqrt{10}\F^{3/2}_1\big)\rra_{t_s} \bigg]\,,\label{eq:PWI12} 
\end{align}
where
\begin{align}
 A_x&=\frac{\kappa(x)}{2x}\,, & B_x&=\frac{\kappa(x)(x^2+\Delta)}{2x^2}\,,\nnnl
 C_x&=\frac{(x^2-\Delta)^2-x^2(\Sigma_0-2x)^2}{4x^2}\,, & D_x&=-\frac{3x^2-\Delta-x\Sigma_0}{2x}\,,   \label{eq:ABCDx}
\end{align}
with $\Sigma_0 = M_D^2+M_K^2+2M_\pi^2$,  $x\in\{s,t\}$.
Thus from Eq.~\eqref{eq:PWI12}, the inhomogeneity can be immediately read off from the 
relation $\sqrt{3}\M_{-++}^{1/2,0}(s)=\F^{1/2}_0(s)+\hat{\F}^{1/2}_0(s)$.
The full set of inhomogeneities is given in terms of the angular averages
{\allowdisplaybreaks
\begin{align}
\hat{\F}^2_0(u)=&\frac{2}{\sqrt{3}}\bigg[\Big\la \F^{1/2}_0-\sqrt{\frac{2}{5}}\F^{3/2}_0\Big\ra_{s_u}\nnnl
 &-\Big\la\big(A_u z^2-B_u z- C_u\big)z^2\bigg(\F^{1/2}_1-\sqrt{\frac{2}{5}}\F^{3/2}_1\bigg)\Big\ra_{s_u}\bigg]\,,\nnnl
\hat{\F}^1_1(u)=&\frac{2}{\kappa_u(u)}\bigg[\lla z\big(\F^{1/2}_0+\sqrt{10}\F^{3/2}_0\big)\rra_{s_u}\nnnl
&-\lla \big(A_u z^3-B_u z^2 -C_u z\big)\big(\F^{1/2}_1+\sqrt{10}\F^{3/2}_1\big)\rra_{s_u}\bigg]\,,\nnnl
\hat{\F}^{1/2}_0(s)=&\frac{5\sqrt{3}}{6}\lla\F^2_0\rra_{u_s}
+\frac{1}{2}\lla \big(A_s z+D_s\big)\,\F^1_1\rra_{u_s}\nnnl
&+\frac{1}{3}\bigg[\lla 2\F^{1/2}_0-\sqrt{10}\F^{3/2}_0\rra_{t_s}
+\lla \big(A_s^2 z^2+B_s z+ C_s\big) \big(2\F^{1/2}_1-\sqrt{10}\F^{3/2}_1\big)\rra_{t_s} \bigg]\,,\nnnl
\hat{\F}^{1/2}_1(s)=& \frac{1}{\kappa(s)}\bigg[\frac{5\sqrt{3}}{2}\lla z \F^2_0 \rra_{u_s}+\frac{3}{2}\lla \big(A_s z^2+D_s z \big)\,\F^1_1\rra_{u_s}\nnnl
&+\lla 2z\F^{1/2}_0-\sqrt{10}z\F^{3/2}_0\rra_{t_s}
+\lla \big(A_s^2 z^3+B_s z^2+C_2 z\big) \big(2\F^{1/2}_1-\sqrt{10}\F^{3/2}_1\big)\rra_{t_s}\bigg]\,, \nnnl
\hat{\F}^{3/2}_0(s)=&-\frac{\sqrt{5}}{2\sqrt{6}}\lla\F^2_0 \rra_{u_s}
+\frac{\sqrt{5}}{2\sqrt{2}}\lla \big(A_s z+D_s\big)\,\F^1_1\rra_{u_s}\nnnl
&-\frac{1}{6}\bigg[\lla\sqrt{10}\F^{1/2}_0+4\F^{3/2}_0\rra_{t_s}
+\lla \big(A_s^2 z^2+B_s z +C_s\big) \big(\sqrt{10}\F^{1/2}_1+4\F^{3/2}_1\big)\rra_{t_s}\bigg]\,. \nnnl
\hat{\F}^{1/2}_2(s)=&\frac{1}{\kappa^2(s)}\bigg[\frac{25}{2\sqrt{3}}\lla (3z^2-1)\,\F^2_0\rra_{u_s}+\frac{5}{2}\lla \big(A_s z+D_s\big) \big(3z^2-1\big)\,\F^1_1\rra_{u_s}\nnnl
&+\frac{10}{3}\lla (3z^2-1)\,\F^{1/2}_0\rra_{t_s}
-\frac{15\sqrt{10}}{3}\lla (3z^2-1)\,\F^{3/2}_0\rra_{t_s}\nnnl
&+\frac{10}{3}\lla \big(A_s^2 z^2 +B_s z+C_s\big)(3z^2-1)\,\F^{1/2}_1\rra_{t_s}\bigg]\,,
\end{align}
}%
where in addition to Eq.~\eqref{eq:ABCDx} we have used
\beq
 A_u=\frac{1}{4}\kappa_u(u)^2\,, \qquad B_u=\frac{1}{2}u\kappa_u(u)\,, \qquad
 C_u=\frac{\left(\Sigma_0-2u\right)^2-u^2}{4}-\Delta\,. 
\eeq

\section{Invariance group matching}\label{app:Invariance}

In this appendix, we study the polynomial ambiguities in the decomposition of the total decay amplitudes 
Eq.~\eqref{eq:Fullamp} into single-variable functions, dubbed ``invariance group''.  
We wish to determine the polynomial at most linear in the Mandelstam variables
that can be added to the different single-variable amplitudes, 
leaving the total decay amplitudes Eq.~\eqref{eq:Fullamp} invariant.
For this purpose, we make use of the relation $s+t+u=3s_0 = M_D^2+M_K^2+2M_\pi^2$.
It is easy to check that adding the following terms to the various $S$-waves as well as the $\pi\pi$ $P$-wave:
\begin{align}
{\F_0^2}^{\text{inv}}(u)&=a_0+b_0u\,, &
{\F_1^1}^{\text{inv}}(u)&=-\frac{5}{\sqrt{3}}b_0+2d_0\,, \nnnl
{\F_0^{1/2}}^{\text{inv}}(s)&=c_0+d_0s\,,&
{\F_0^{3/2}}^{\text{inv}}(s)&=\frac{\sqrt{5}}{2\sqrt{2}}\left(\sqrt{3}\big[a_0+b_0(3s_0-2s)\big]+2(c_0+d_0s)\right)\,,
\label{eq:invgroup}
\end{align}
leaves both $\M_{-++}(s,t,u)$ and $\M_{\ol{0}0+}(s,t,u)$ unchanged.
The most general full decay amplitudes are therefore obtained by 
 \beq
 {\F_L^I}^{\text{new}}(s)=\F_L^I(s)+{\F_L^I}^{\text{inv}}(s)\,,
 \eeq
which, according to Eq.~\eqref{eq:invgroup}, has a four-parameter gauge freedom built in.

Following Ref.~\cite{Leutwyler}, we rewrite the polynomial representations of ${\F_L^I}^{\text{inv}}(s)$ Eq.~\eqref{eq:invgroup} into the Omn\`es representation ${\F_L^I}^{\text{inv}}_{\Omega}(s)$ in order to match to Eq.~\eqref{eq:Khuri}:
\beq
 {\F_L^I}^{\text{inv}}_{\Omega}(s)\equiv \Omega_L^I(s)\bigg\{\pi_L^I(s)+\frac{s^n}{\pi}\int_{s_{\rm th}}^\infty\frac{\diff x}{x^n}~\frac{\sin \delta_L^I(x){\hat{\F}_L^{I^{\text{inv}}}}}{|\Omega_L^I(x)|(x-s)}\bigg\}\,,
\eeq
with the subtraction polynomials  $\pi_L^I(s)$.
As the invariance polynomials ${\F_L^I}^{\text{inv}}(s)$ do not have discontinuities, it immediately follows that
${{\hat{\F}}_L^{I^{\text{inv}}}}(s)=-{\F_L^I}^{\text{inv}}(s)$, which is also confirmed by a straightforward calculation.
We determine the subtraction polynomials by equating the polynomial and Omn\`es representations of the invariance group. We obtain 
\beq \label{eq:invmatch}
\pi_L^I(s)=\frac{{\F_L^I}^{\text{inv}}(s)}{\Omega_L^I}+\frac{s^n}{\pi}\int_{s_{\rm th}}^\infty\frac{\diff x}{x^n}~\frac{\sin \delta_L^I(x){{\F}_L^I}^{\text{inv}}}{|\Omega_L^I(x)|(x-s)}\,.
\eeq
The next step is to rewrite the inverse Omn\`es function into a dispersion relation.
Its discontinuity is given by
 \beq
 \text{disc}\frac{1}{\Omega_L^I(s)}=-2i \frac{\sin \delta_L^I(x)}{|\Omega_L^I(x)|}\,,
 \eeq
 which thus yields
 \beq
 \frac{1}{\Omega_L^I(s)}={P_L^I}_\Omega(s)-\frac{s^n}{\pi}\int_{s_{\rm th}}^\infty\frac{\diff x}{x^n}~\frac{\sin \delta_L^I(x)}{|\Omega_L^I(x)|(x-s)}\,,
 \eeq
 with the subtraction polynomial ${P_L^I}_\Omega(s)=1+\sum_{i=1}^{n-1}(\omega_L^I)_i s^i$.
 The subtraction constants $(\omega_L^I)_i$ are given by the following sum rules, provided that the dispersion integrals converge:
 \beq
(\omega_L^I)_i=-\frac{1}{\pi}\int_{s_{\rm th}}^\infty \frac{\diff x}{x^{i+1}}\frac{\sin \delta_L^I(x)}{|\Omega_L^I(x)|}\,.
 \eeq
 Therefore Eq.~\eqref{eq:invmatch} yields
 \beq
\pi_L^I(s)={P_L^I}_\Omega(s){{\F}_L^I}^{\text{inv}}(s)+\frac{s^n}{\pi}\int_{s_{\rm th}}^\infty\frac{\diff x}{x^n}~\frac{\sin \delta_L^I(x)\big({\F}_L^{I^{\text{inv}}}(x)-{\F}_L^I{^{\text{inv}}}(s)\big)}{|\Omega_L^I(x)|(x-s)}\,.
 \eeq
 As an example we will study the single-variable amplitude $\F^2_0$ with ${\F}_0^{2^{\text{inv}}}(s)=a_0+b_0s$. We obtain
 \beq
 \pi_0^2(s)=a_0+\Big[b_0+a_0(\omega_0^2)_1\Big]s+\left((\omega_0^2)_1-\frac{1}{\pi}\int_{s_{\rm th}}^\infty\frac{\diff x}{x^2}~\frac{\sin \delta_L^I(x)}{|\Omega_L^I(x)|}\right)b_0s^2\,.
 \eeq
 Using the sum rule value for $(\omega_0^2)_1$ we find
 \beq
 \pi_0^2(s)=a_0+\Big[b_0+a_0(\omega_0^2)_1\Big]s \,.
 \eeq
 The other subtraction polynomials are  obtained in an analogous way and read
\begin{align}
\pi_0^2(s)&=a_0+(b_0+a_0(\omega_0^2)_1)s\, , \qquad
\pi_1^1(s)=-\frac{5}{\sqrt{3}}b_0+2d_0 \, ,  \nnnl
\pi_0^{1/2}(s)&=c_0+\Big[d_0+(\omega_0^{1/2})_1\Big]s+\Big[d_0(\omega_0^{1/2})_1+c_0(\omega_0^{1/2})_2\Big]s^2+\Big[d_0(\omega_0^{1/2})_2+c_0(\omega_0^{1/2})_3\Big]s^3\,  \nnnl
\pi_0^{3/2}(s)&=\frac{\sqrt{5}}{2\sqrt{2}}\bigg\{ \sqrt{3}(a_0+3b_0s_0)+2c_0  \nnnl
 & \qquad \qquad
+\Big[(\omega_0^{3/2})_1 \left(\sqrt{3} (a_0+3b_0s_0)+2c_0\right)- 2\left(\sqrt{3}b_0-d_0\right)\Big]s \bigg\}\,, \label{eq:gaugepoly}
\end{align}
with no contributions to the $\pi K$ $P$- and $D$-waves.
Polynomial terms with higher order than the subtraction polynomials of the corresponding amplitudes (see Sec.~\ref{subsec:subtraction}) have been omitted.

As we have argued above that a choice of the constants $a_0$, \ldots, $d_0$ corresponds to a mere ``gauge'' choice
and is unobservable, we can decide to fix them by requiring the (linear) subtraction polynomials in the nonresonant
$S$-waves ($I=2$ $\pi\pi$ and $I=3/2$ $\pi K$) to vanish.  Equation~\eqref{eq:gaugepoly} proves that this is feasible:
we can eliminate all subtraction constants in $\F_0^2$ by the appropriate choice of $a_0$ and $b_0$, 
and all constants in $\F_0^{3/2}$ by adjusting $c_0$ and $d_0$.  The result is the system 
Eq.~\eqref{eq:fulleq} in the main text, which is thus free of ambiguities.

\section{Numerical treatment}\label{app:Numerics}
In this appendix we discuss the numerical treatment of the double integral in Eq.~\eqref{eq:Fint}, i.e.\
the part independent of the subtraction constants.
In the following we will restrict ourselves to the $s$-channel case for illustration.
We rewrite the term $z_s^m$ in Eq.~\eqref{eq:Fint} as  $z_s^m(s,t)=\zeta^m(s,t)/\kappa^m(s)$, with $\zeta(s,t)=\big(2ts-3s_0s+s^2-\Delta\big)$ and $3s_0 = M_D^2+M_K^2+2M_\pi^2$, such that the double integral adopts the form

\beq
\tilde{\F}(s)=s\,\kappa^{2L-m}(s)\int_{s_{\rm th}}^\infty\frac{\tilde{\F}(x)\sin\delta(x)}{x^n|\Omega(x)|\kappa^{2L+1}(x)}\int_{t_-(s)}^{t_+(s)}t^n\frac{\zeta^m(s,t)\Omega(t)}{x-t}\,\diff t\,\diff x\,,
\eeq
with $t_\pm(s)=t(s,\pm1)$.
First we study the case $s>\left(\mD-\mpi\right)^2$.
The angular integral can directly be performed as the two integral paths do not cross each other. We may simply use
\beq
s\,\kappa^{2L-m}(s)\int_{s_{\rm th}}^\infty\frac{\tilde{\F}(x)\sin\delta(x)}{x^n|\Omega(x)|\kappa^{2L+1}(x)}W(s,x)\,\diff x\,, \quad
W(s,x)\equiv \int_{t_-(s)}^{t_+(s)}t^n\frac{\zeta^m(s,t)\Omega(t)}{x-t}\,\diff t\,,
\eeq
where $W(s,x)$ can be determined numerically in a straightforward way.
The discretized integral reads
\beq\label{eq:numdisc}
\int_{s_{\rm th}}^\infty\frac{\tilde{\F}(x)\sin\delta(x)}{x^n|\Omega(x)|\kappa^{2L+1}(x)}W(s,x)\,\diff x=\sum_j \tilde{\F}(s_j)\int_{s_j}^{s_{j+1}} \frac{c_0^j(s)+c_1^j(s)x}{\kappa^{2L+1}(x)}\,\diff x\,, 
\eeq
where $c_0^j(s)+c_1^j(s)x$ is the linear interpolation of ${W(s,x)\sin\delta(x)}/{x^n|\Omega(x)|}$ in the interval $[s_j,s_{j+1}]$ for a fixed $s$.
Note that the resulting integrals can be performed analytically with the singularities moved into the upper complex plane to obtain the correct (physical) branch.

For the case $s<\left(\mD-\mpi\right)^2$ the Cauchy singularity needs to be handled carefully,
 as the integration paths meet.  We rewrite
\begin{align}
&\int_{s_{\rm th}}^\infty\tilde{\F}(x)\frac{\sin\delta(x)}{x^n|\Omega(x)|\kappa^{2L+1}(x)}\int_{t_-(s)}^{t_+(s)}t^n\frac{\zeta^m(s,t)\Omega(t)}{x-t}\,\diff t\,\diff x \nnnl
=&\int_{s_{\rm th}}^\infty\frac{\tilde{\F}(x)\sin\delta(x)}{x^n|\Omega(x)|\kappa^{2L+1}(x)}\int_{t_-(s)}^{t_+(s)}\zeta^m(s,t)\frac{t^n\Omega(t)- x^n\Omega(x)}{x-t}\,\diff t\,\diff x \nnnl
  &+\int_{s_{\rm th}}^\infty\int_{t_-(s)}^{t_+(s)}\frac{\tilde{\F}(x)e^{i\delta(x)}\sin\delta(x)}{\kappa(x)^{2L+1}(x-t)}\,\diff t\,\diff x\,.
\end{align}
The first summand is treated as above in Eq.~\eqref{eq:numdisc}. For the second summand we obtain 
\beq
\int_{s_{\rm th}}^\infty\int_{t_-(s)}^{t_+(s)}\frac{\tilde{\F}(x)e^{i\delta(x)}\sin\delta(x)}{\kappa^{2L+1}(x)(x-t)}\,\diff t\,\diff x
=\sum_j \tilde{\F}(s_j)\int_{s_j}^{s_{j+1}}\int_{t_-(s)}^{t_+(s)}\frac{a_0^j+a_1^jx}{\kappa^{2L+1}(x)(x-t)}\,\diff t\,\diff x\,,
\eeq
where now $a_0^j+a_1^jx$ is the linear interpolation of $e^{i\delta(x)}\sin\delta(x)$ in the interval $[s_j,s_{j+1}]$.

\end{document}